# Actinium hydrides $AcH_{10}$, $AcH_{12}$, $AcH_{16}$ as high-temperature conventional superconductors

*Dmitrii V. Semenok*,* [1,2] *Alexander G. Kvashnin,* [1,2] *Ivan A. Kruglov,* [2,3] *Artem R. Oganov*,* [1,2,3,4]

[1] Skolkovo Institute of Science and Technology, Skolkovo Innovation Center 143026, 3 Nobel Street, Moscow, Russian Federation

[2] Moscow Institute of Physics and Technology, 141700, 9 Institutsky lane, Dolgoprudny, Russian Federation

[3] Dukhov Research Institute of Automatics (VNIIA), Moscow 127055, Russian Federation

[4] International Center for Materials Discovery, Northwestern Polytechnical University, Xi'an, 710072, China

**Corresponding Author**

*Dmitrii V. Semenok, dmitrii.semenok@skolkovotech.ru

*Artem R. Oganov, A.Oganov@skoltech.ru

**ABSTRACT.** Stability of numerous unexpected actinium hydrides was predicted *via* evolutionary algorithm USPEX. Electron-phonon interaction was investigated for the hydrogen-richest and most symmetric phases: $R\bar{3}m$-$AcH_{10}$, *I4/mmm*-$AcH_{12}$ and $P\bar{6}m2$-$AcH_{16}$. Predicted structures of actinium hydrides are consistent with all previously studied Ac-H phases and demonstrate phonon-mediated high-temperature superconductivity with $T_C$ in the range 204-251 K for $R\bar{3}m$-$AcH_{10}$ at 200 GPa and 199-241 K for $P\bar{6}m2$-$AcH_{16}$ at 150 GPa which was estimated by directly solving of Eliashberg equation. Actinium belongs to the series of $d^1$-elements (Sc-Y-La-Ac) that form high-$T_C$ superconducting (HTSC) hydrides. Combining this observation with $p^0$-HTSC hydrides ($MgH_6$ and $CaH_6$), we propose that $p^0$- and $d^1$-atoms with low-lying empty orbitals tend to form phonon-mediated HTSC metal polyhydrides.

**Graphical Abstract**

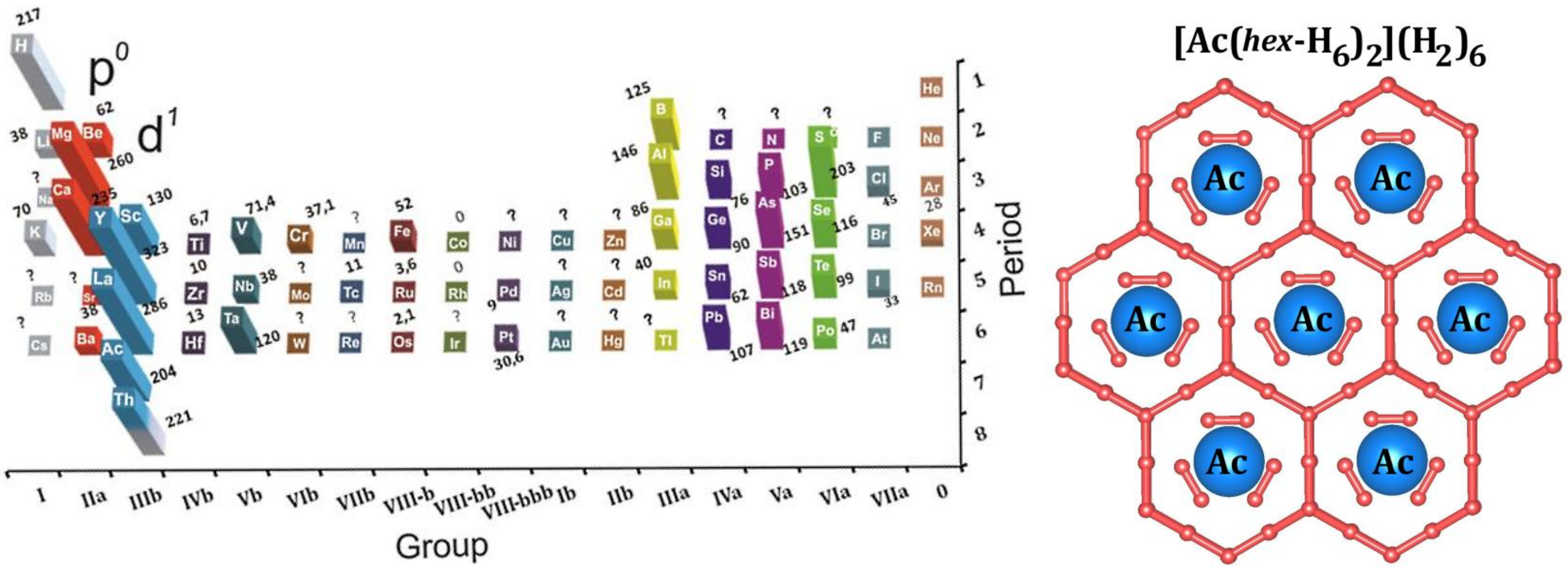

## Introduction

Computational search for phonon-mediated superconductivity in binary hydrides is coming to the finale. Recent results, presented on the Mendeleev's table on Figure S1 (see Supporting Information) show that *p*-elements from IIIa-VIa groups form superconducting hydrides with an evenly distributed critical temperature that generally decreases with increasing period number and atomic mass of hydride-forming atom. A completely different situation takes place for *s, d, f*-metals which form a pronounced belt of superconductivity (groups IIa-IIIb, Figure S1). Recent predictions of potential high-temperature superconductivity in lanthanum and yttrium hydrides ($LaH_{10}$ and $YH_{10}$ [1]), and hydrides of uranium ($UH_8$ [2]), thorium ($ThH_{10}$ [3]), scandium ($ScH_6$ [4]) hydrides have one thing in common: all these elements contain almost empty *d*, *f*-shells or *p*-shells, ($MgH_6$, $CaH_6$) and form what we call a "lability belt" (see Table 1), i.e. a zone of elements with electronic structure particularly sensitive to external conditions, because they have empty low-lying orbitals the population of which depends on the atomic environment.

**Table 1.** The most promising predicted superconducting hydrides from so-called "lability belt"

| Hydride | Electronic configuration of the atom | Calculated max $T_C$ (pressure, GPa) |
|---|---|---|
| $ScH_6$ | [Ar] **3d¹** 4s² | 130 K (200) |
| $YH_{10}$ | [Kr] **4d¹** 5s² | 326 K (300) |
| $LaH_{10}$* | [Xe] **5d¹** 6s² | 286 K (250) |
| $AcH_{10}$ | [Rn] **6d¹** 7s² | 251 K (this work) |
| $ThH_{10}$ | [Rn] **6d²** 7s² | 221 K (100) |
| *$UH_8$* | *[Rn] 5f³ **6d¹** 7s²* | *193 K (0)* |

*Structure has been confirmed experimentally [5].

Sequential occupation of the electronic orbitals, for example, in the lanthanide series, leads to a systematic decrease of the electron-phonon coupling (EPC) strength [6]. A possible explanation is that under high pressure electron shells with different symmetry come closer in energy and H-atoms become able to form electron-deficient (e.g., 3c2e [7]) bonds with a central atom using a large number of vacant *p, d* or *f*-orbitals. Increasing the number of electrons on valence orbitals of the central metal atom leads an increase of the Coulomb electron-electron repulsion that sharply reduces the electron-phonon interaction.

Another factor should also be taken into account. For $s^1$-$s^2$ metals from I-IIa groups (e.g. K, Ca, Sr, Ba) the empty *d*-shells are quite close in energy to the *s*-shell. It means that small structural changes in the environment of central atom may lead to large relative changes in the energy and occupation of *d*-orbitals. This "lability factor" may help us to understand an unexpected increase of electron-phonon interaction in hydrides of boundary (*s↔p, s↔d, d↔f*) metals.

Verification of this hypothesis requires the study of actinide hydrides, which have sufficiently diverse structure of electronic shells. Direct analogy of chemical properties, atomic size, electronegativity and electronic configuration of actinium and lanthanum and recent experimental synthesis of $LaH_{10+x}$ [5] along with predictions of near-room temperature superconductivity in $LaH_{10}$ inspired us to study the phase diagram and possible superconductivity in the Ac-H system.

Actinium (from the Greek ἀκτίς – shining) is highly radioactive; its longest-lived isotope $^{227}$Ac has half-life of 22 years and in the bulk phase has fcc crystal structure. Actinium is chemically active and in its properties (valence is 3, atomic radius is 1.88 Å) is similar to lanthanum. Isotope $^{227}$Ac is usually obtained by irradiation of radium with neutrons (or deuterons) in the reactor with a small yield ~ 2.15 %. Radioactive decay of $^{227}$Ac predominantly proceeds as $\beta^-$ decay with the formation of $^{227}$Th and further $^{223}$Ra through the α-decay. [8] Thus, due to radioactivity of daughter nuclides $^{227}$Ac is active as α-emitter found applications in Ac-Be neutron sources, [9] thermoelectric generators [10] and in nuclear medicine for experimental treatment of some forms of cancer, e.g. myeloid leukemia [11]. Recently, another exciting property of Ac – formation of polycations $AcHe_{17}^{3+}$ with helium, was predicted [12]. This is closely related to a synthesis and study of polyhydrides (e.g. $XH_{16}$) as a particular case of polyatomic complexes and its stabilization by pressure or charge.

Superconductivity of metallic actinium under pressure was studied theoretically by Dakshinamoorthy et al. [13],where the critical transition temperature ($T_C$) to superconducting state was predicted to be from 5 to 12 K. It was also theoretically found that actinium hydride $AcH_2$ is metallic and undergoes multiple $Fm\bar{3}m \rightarrow P4_2/mmc \rightarrow Imma \rightarrow P6_3/mmm$ phase transitions at 12, 27 and 68 GPa, respectively. [14] Due to the high cost and radioactivity of actinium, its chemistry has been studied very poorly. In this paper, we intend to partially fill this gap.

## Computational Methodology

Evolutionary algorithm USPEX [15–17] is a powerful tool for prediction of thermodynamically stable compounds of given elements at a given pressure (for some of its application, see Refs. [18–20]). We performed variable-composition searches in the Ac-H system at pressures 0, 50, 150, 250 and 350 GPa. The first generation (110 structures) was created using random symmetric generator, while all subsequent generations contained 20% random structures, and 80% created using heredity, softmutation and transmutation operators.

Here, evolutionary investigations were combined with structure relaxations and total energy calculations using density functional theory (DFT) [21,22] within the generalized gradient approximation (Perdew-Burke-Ernzerhof functional), [23] and the projector augmented wave method [24,25], as implemented in the VASP code. [26–28] Plane wave kinetic energy cutoff was set to 600 eV and the Brillouin zone was samples by Γ-centered k-points meshes with resolution $2\pi\times0.05$ Å$^{-1}$.

In order to establish stability fields of the predicted phases, we recalculated their enthalpies with increased precision at various pressures with a smaller pressure increment (from 5 to 10 GPa), recalculating the thermodynamic convex hull (Maxwell construction) at each pressure. By definition, the thermodynamically stable phase has lower Gibbs free energy (or, at zero Kelvin, lower enthalpy) than any phase or phase assemblage of the same composition. Phases that were located on the convex hull are the ones stable at given pressure. Stable structures of elemental Ac, H and $AcH_3$ were taken from USPEX calculations and agree with those from Refs. [29,30].

Calculations of superconducting $T_C$ were carried out using QUANTUM ESPRESSO package [31]. Phonon frequencies and electron-phonon coupling (EPC) coefficients were computed using density-functional perturbation theory [32], employing plane-wave pseudopotential method and Perdew-Burke-Ernzerhof exchange-correlation functional [23]. The effects of core electrons Ac on valence wave functions were modeled with the aid of scalar relativistic ultrasoft pseudopotentials [33]. The $6s^2p^6d^17s^2$ electrons of Ac were treated explicitly. Convergence tests showed that 80 Ry is a suitable kinetic energy cutoff for the plane wave basis set.

Electronic band structures of newly found actinium hydrides were calculated using both VASP and QE, and demonstrated good consistency. Comparison of phonon densities of states calculated using finite displacement method (VASP and PHONOPY [34,35]) and density-functional perturbation theory (QE) showed perfect agreement between these methods.

Critical temperature was calculated from the Eliashberg equation [36] which is based on the Fröhlich Hamiltonian $\hat{H} = \hat{H}_e + \hat{H}_{ph} + \sum_{k,q,j} g_{k+q,k}^{q,j} \hat{c}_{k+q}^{+} \hat{c}_k (\hat{b}_{-q,j}^{+} + \hat{b}_{q,j})$, where $c^+$, $b^+$ relate to creation operators of electrons and phonons, respectively. Matrix element of electron-phonon interaction $g_{k+q,k}^{q,j}$ calculated within the harmonic approximation in Quantum ESPRESSO can be defined as $g_{k+q,k}^{q,j} = \sqrt{\frac{\hbar}{2M\omega_{q,j}}} \int \psi_k^*(r) \cdot \left\{ \frac{dV_{scf}}{d\vec{u}_q} \cdot \frac{\vec{u}_q}{|\vec{u}_q|} \right\} \cdot \psi_{k+q}(r) d^3 r$, where $u_q$ is the displacement of an atom with mass $M$ in the phonon mode $q,j$. Within the framework of Gor'kov and Migdal approach [37,38] the correction to the electron Green's function $\Sigma(\vec{k},\omega) = G_0^{-1}(\vec{k},\omega) - G^{-1}(\vec{k},\omega)$ caused by interaction can be calculated by taking into account only the first terms of the expansion of electron-phonon interaction in series of $(\omega_{log}/E_F)$. As a result, it will lead to integral Eliashberg equations [36]. These equations can be solved by iterative self-consistent method for the real part of the order parameter Δ(T, ω) (superconducting gap) and the mass renormalization function Z(T, ω) [39] (see Supporting Information).

In our *ab initio* calculations of the electron-phonon coupling (EPC) parameter λ, the first Brillouin zone was sampled using 4×4×4 q-points mesh, and a denser 24×24×24 k-points mesh (with Gaussian smearing and σ = 0.02 Ry, which approximates the zero-width limits in the calculation of λ). The superconducting transition temperature $T_C$ was also estimated by using two equations: "full" – Allen-Dynes and "short" – modified McMillan equation [40]. The "full" Allen-Dynes equation for calculating $T_C$ has the following form [40]:

$$T_C = \omega_{log} \frac{f_1 f_2}{1.2} \exp\left( \frac{-1.04(1+\lambda)}{\lambda - \mu^* - 0.62\lambda\mu^*} \right) \quad (1)$$

with

$$f_1 f_2 = \sqrt[3]{1 + \left( \frac{\lambda}{2.46(1+3.8\mu^*)} \right)^{\frac{3}{2}}} \cdot \left( 1 - \frac{\lambda^2 (1 - \omega_2/\omega_{log})}{\lambda^2 + 3.312(1+6.3\mu^*)^2} \right), \quad (2)$$

while the modified McMillan equation has the form but with $f_1 * f_2 = 1$.

The EPC constant λ, logarithmic average frequency $\omega_{log}$ and mean square frequency $\omega_2$ were calculated as:

$$\lambda = \int_0^{\omega_{max}} \frac{2 \cdot a^2 F(\omega)}{\omega} d\omega \quad (3)$$

and

$$\omega_{log} = \exp\left( \frac{2}{\lambda} \int_0^{\omega_{max}} \frac{d\omega}{\omega} \alpha^2 F(\omega) \ln(\omega) \right), \omega_2 = \sqrt{\frac{1}{\lambda} \int_0^{\omega_{max}} \left[ \frac{2a^2 F(\omega)}{\omega} \right] \omega^2 d\omega} \quad (4)$$

where μ* is the Coulomb pseudopotential, for which we used widely accepted lower and upper bounds of 0.10 and 0.15.

## Results

We built the composition-pressure phase diagram (see Figure 1), which shows pressure ranges of stability of all phases found at different pressures. As shown in Figure 1, our calculations correctly reproduce stability of $Fm\bar{3}m$-$AcH_3$ [29] and predict 12 new stable phases. Due to the fact that conditionally stable (extremely low-energy metastable) phases can be found in experiment, or even more can dominate over thermodynamically stable phases (e.g., Ref. [41]), we extended the stability region for $R\bar{3}m$-$AcH_{10}$ and *I4/mmm*-$AcH_{12}$ to higher pressures shown in Figure 1 by the yellow gradient and dotted lines.

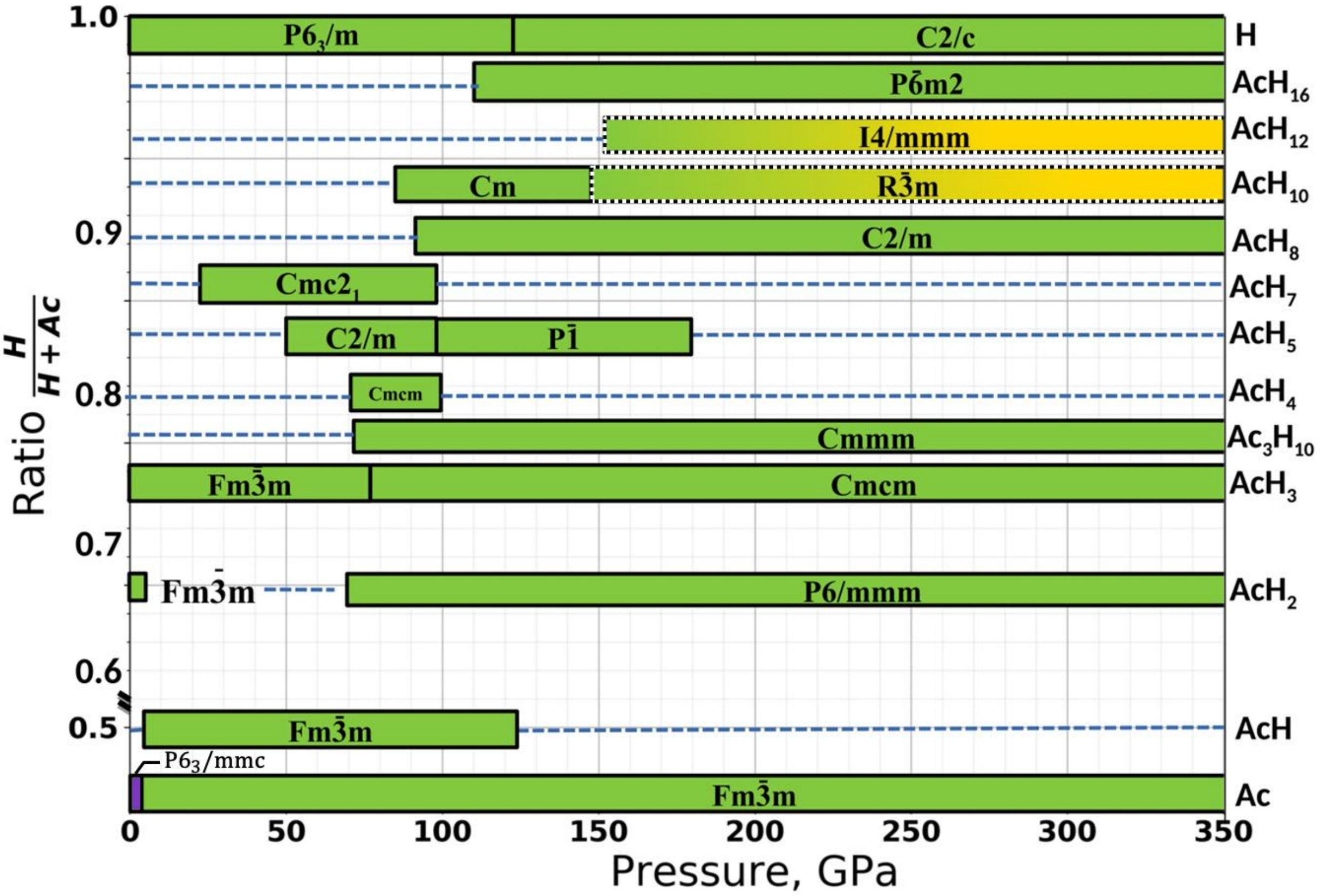

Figure 1. Pressure-composition phase diagram of the Ac-H system. Dotted lines with yellow gradient show pressure ranges where $R\bar{3}m$-$AcH_{10}$ and *I4/mmm*-$AcH_{12}$ may be metastable.

Detailed information on crystal structures of the predicted phases is summarized in Table S2 (see Supporting Information). Our calculations show that at zero pressure elemental Ac has *P6₃/mmc* (α-Ac) structure which is lower in energy than $Fm\bar{3}m$-Ac (β-Ac) by 11 meV/atom and transforms to $Fm\bar{3}m$ phase at 3 GPa. At 0 GPa there are only two stable hydrides – metallic $Fm\bar{3}m$-$AcH_2$ and semiconducting $Fm\bar{3}m$-$AcH_3$ (DFT bandgap is 0.95 eV). $Fm\bar{3}m$-$AcH_2$ has fluorite-type structure with *fcc*-sublattice of actinium atoms, in which hydrogen atoms occupy all tetrahedral voids. Structure of $Fm\bar{3}m$-$AcH_3$ contains an extra H-atom in the octahedral void and belongs to the $Fe_3Al$ structure type. Phase transitions of actinium dihydride ($Fm\bar{3}m \rightarrow P4_2/mmc \rightarrow Imma \rightarrow P6_3/mmm$) predicted in Ref. [14] turned out to be metastable due to decomposition $2AcH_2 \rightarrow AcH + AcH_3$ in the pressure range 5-65 GPa (see Supporting Information).

$Fm\overline{3}m$-$AcH_3$ loses its stability at 78 GPa, when it transforms to *Cmcm*-$AcH_3$, which remains stable at least up to 350 GPa (see Figure 1). In this phase Ac atoms are coordinated by 14 hydrogens (see Supporting Information). Apart from $AcH_2$ and $AcH_3$, at low pressures (>5 GPa) there is another phase $Fm\overline{3}m$-AcH with rocksalt crystal structure, which decomposes at 125 GPa (2AcH → Ac + *P6/mmm*-$AcH_2$).

At 25 GPa a new superhydride *Cmc2₁*-$AcH_7$ appears and remains stable up to 100 GPa. In this structure Ac atoms are surrounded by 10 hydrogen atoms ($d_{Ac\text{-}H}$ = 2.15-2.18 Å at 100 GPa), 4 of them are terminal atoms in stretched $H_2$ molecules ($d_{H\text{-}H}$ = 0.84 Å compared with $d_{H\text{-}H}$ = 0.74 Å in the isolated $H_2$ molecule) while other 6 atoms are single hydrogens. The maximum distance $d_{Ac\text{-}H}$ is 2.3 Å. At 50 GPa *C2/m*-$AcH_5$ becomes stable (see Figure 1), where Ac atoms are connected with 6 H-atoms ($d_{Ac\text{-}H}$ = 2.22 Å at 70 GPa) and two stretched $H_2$ molecules ($d_{H\text{-}H}$ = 0.85 Å at 70 GPa). At 100 GPa this structure undergoes a phase transition to $P\overline{1}$-$AcH_5$, which loses stability at 180 GPa (see Figure 1, 2).

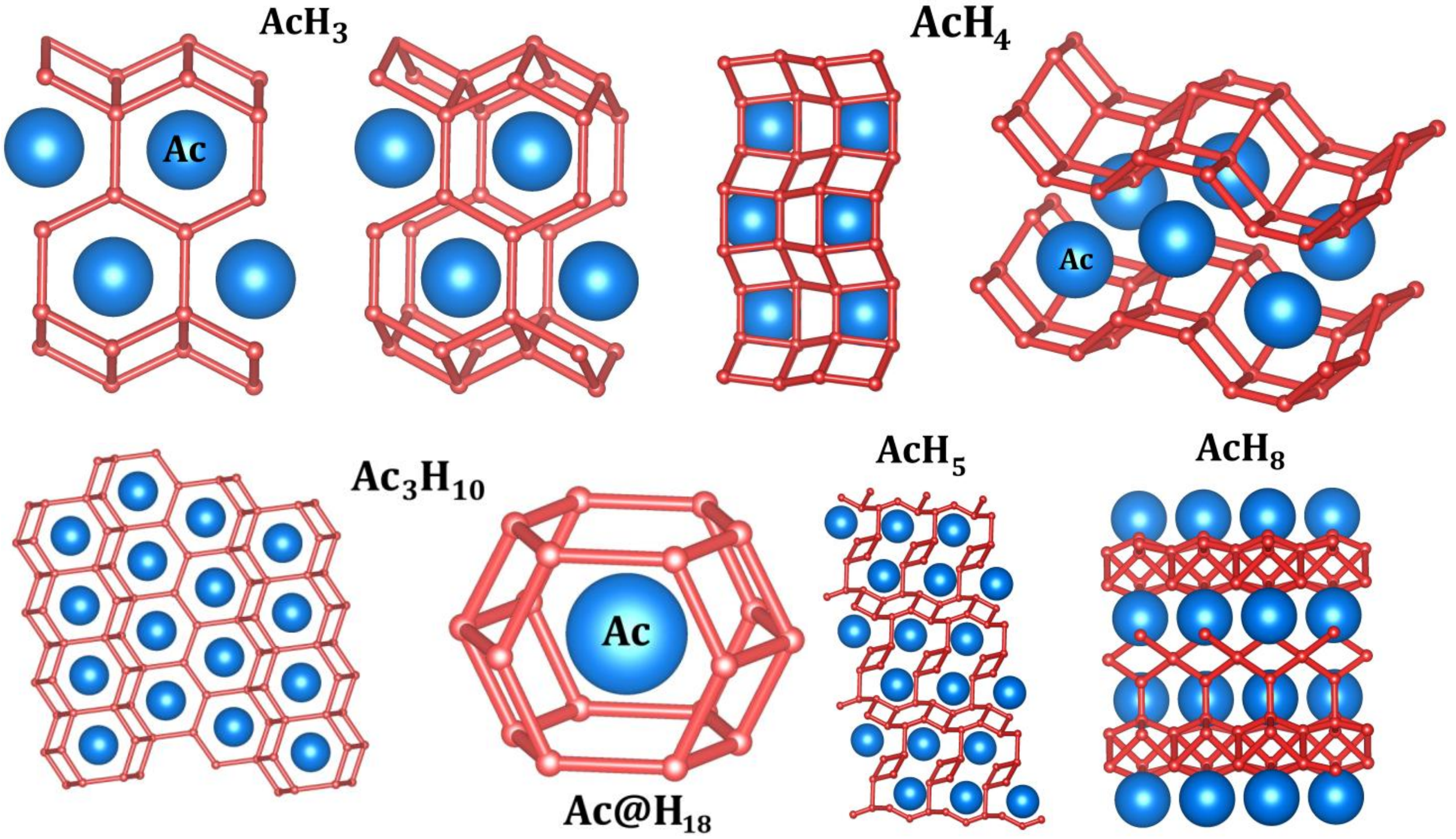

Figure 2. Crystal structures of predicted *Cmcm*-$AcH_3$ (150 GPa, min $d_{HH}$ =1.53 Å), *Cmcm*-$AcH_4$ (100 GPa, min $d_{HH}$ =1.31 Å in $H_2$-fragments), *Cmmm*-$Ac_3H_{10}$ (150 GPa, min $d_{HH}$= 1.58 Å), $P\overline{1}$-$AcH_5$ (150 GPa, min d(H-H)= 0.99 Å in $H_2$ fragments) and $C2/m$-$AcH_8$ (150 GPa, min d(H-H)= 1.06 Å in $H_3$ fragments) phases.

At 75 GPa two new phases *Cmmm*-$Ac_3H_{10}$ and *Cmcm*-$AcH_4$ become stable (see Figure 1). *Cmcm*-$AcH_4$ is stable in a narrow pressure range 75-100 GPa, whereas *Cmmm*-$Ac_3H_{10}$ structurally similar to *Immm*-$Th_3H_{10}$ [3] and *Immm*-$U_3H_{10}$ [2], turns to be stable at least up to 350 GPa (Figure 1). Structurally similar phase *Cmcm*-$AcH_3$ differs by a small displacement of hydrogen atoms (Figure 2), which indicates the possibility of an easy and reversible transformation of $AcH_3$ to $Ac_3H_{10}$ *via* direct hydrogen saturation under pressure.

New metallic hydride *C2/m*-$AcH_8$ becomes stable at 90 GPa and its structure contains bent $H_3$ groups ($d_{H\text{-}H}$ = 1.06 Å, φ = 122° at 150 GPa), and each Ac atom is bonded to three $H_3$ fragments.

At 80 GPa actinium decahydride (*Cm*-$AcH_{10}$) appears on the phase diagram. Around 140 GPa it transforms to a more symmetric $R\overline{3}m$-$AcH_{10}$. Stability of $R\overline{3}m$-$AcH_{10}$ is determined by the enthalpy of reaction $4AcH_{10} \rightarrow 3AcH_8 + AcH_{16}$ which is $\Delta H$= 0.175 eV at 150 GPa (0 K) and $\Delta H$ = -1.628 eV at 250 GPa (0 K) with the accounted ZPE contribution. With increasing of pressure the influence of ZPE vanishes and the decomposition may be more profitable. However, the Helmholtz free energy ($F_{th}$) contribution shifts equilibrium back to the $AcH_{10}$ ($\Delta H$ = 0.309 eV at 150 GPa and 300 K), thus, area of $R\bar{3}m$-$AcH_{10}$ stability expands at room or higher temperatures.

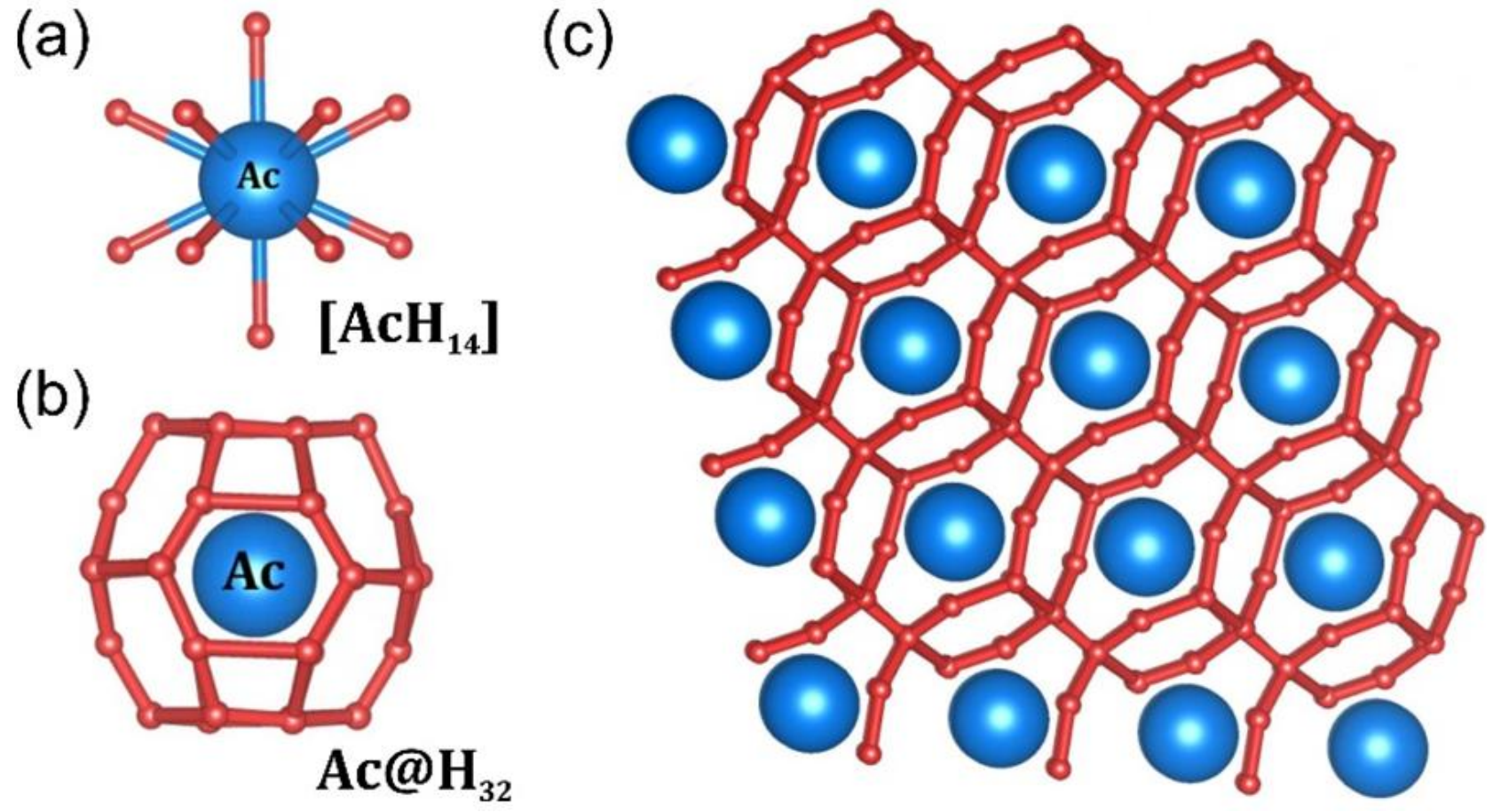

Figure 3. Structural features of $R\bar{3}m$-$AcH_{10}$ at 150 GPa. a) First coordination sphere of Ac atom containing 14 hydrogens; b) hydrogen polyhedral cage around each Ac atom; c) crystal structure of $AcH_{10}$.

In $R\bar{3}m$-$AcH_{10}$ (150 GPa) the first coordination sphere of Ac consists of 14 hydrogen atoms at $d_{Ac\text{-}H}$ = 2.10 Å (see Figure 3a). Hydrogen environment of Ac atoms can be also described as clathrate-like closed shell with 32 hydrogen atoms with diameter ~4.5 Å (see Figure 3b). The shortest H-H distance is 1.07 Å.

The metastable $I4/mmm$-$AcH_{12}$ was studied in the pressure range 150-300 GPa. Existence of this compound is determined by two possible reactions: $3AcH_{12} \rightarrow 2AcH_{10} + AcH_{16}$ and $2AcH_{12} \rightarrow AcH_8 + AcH_{16}$. Calculations indicate that ZPE and $F_{th}$ contributions play here a minor role, and $AcH_{12}$ remains metastable (higher than convex hull by 0.05-0.06 eV/atom).

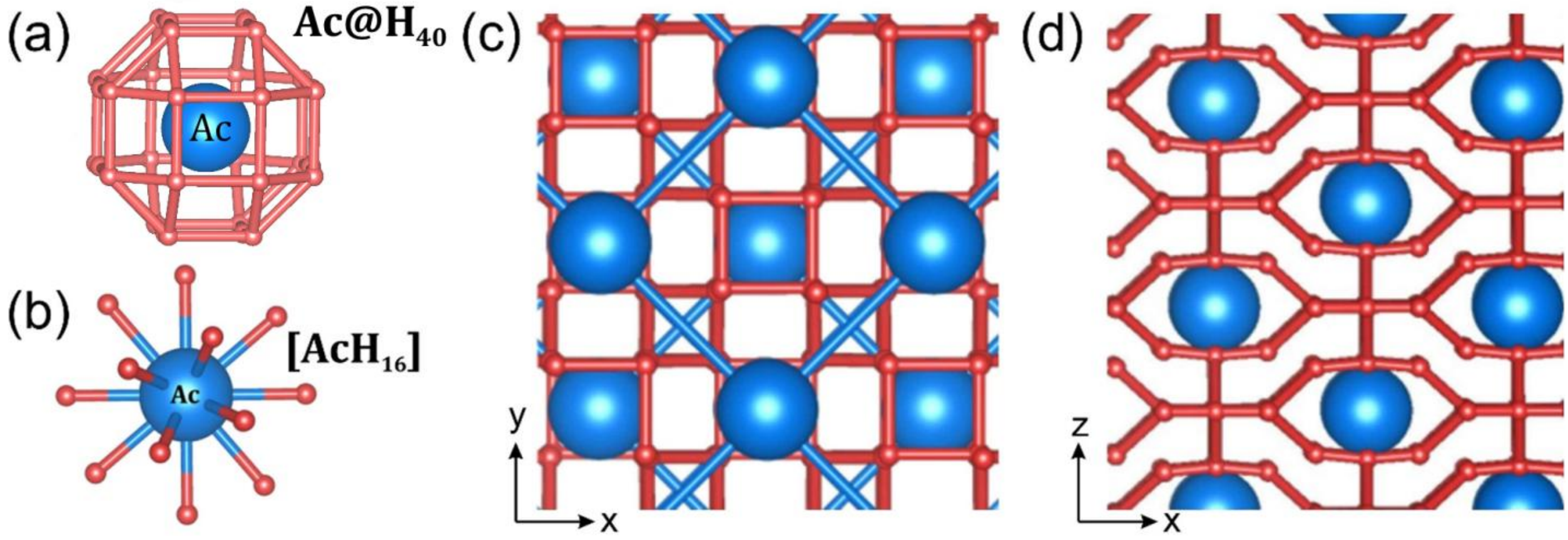

Figure 4. Structural features of $I4/mmm$-$AcH_{12}$ at 150 GPa. a) Coordination sphere of Ac atom containing 16 hydrogen atoms; b) top and c) side views of $I4/mmm$-$AcH_{12}$.

Hydrogen environment of Ac atoms in $I4/mmm$-$AcH_{12}$ can be described as clathrate-like closed shell with 40 hydrogen atoms with diameter ~4.47 Å (see Figure 4a). The coordination sphere of Ac in $I4/mmm$-$AcH_{12}$ is made of 8 H-atoms (see Figure 4b) at $d_{Ac\text{-}H}$ = 2.06 Å and another 8 H-atoms at $d_{Ac-H} = 2.11$ Å at 150 GPa. Connection between [$AcH_{16}$] (Figure 4b) is carried out *via* $H_4$ tetrahedra with $d_{H-H} = 1.45$ Å, but minimum H-H distance in $AcH_{12}$ is in almost linear ($\varphi = 170^o$) $H_3$ fragments with $d_{H-H} = 0.94$ Å (see Figure 4b, c).

An interesting metallic $P\bar{6}m2$-$AcH_{16}$ phase is found to be stable at 110 GPa, in which Ac atoms are coordinated by nearest 12 hydrogens with $d_{Ac-H} = 2.08$ Å at 150 GPa (see Figure 5a). Extension of the radius of coordination sphere up to 2.16 Å leads to addition of 12 more H-atoms as the external coordination sphere (Figure 5b). These 12 H-atoms form 6 $H_2$ molecules ($d_{H-H} = 0.87$ Å) which fill the space between Ac-layers separated by 3.77 Å at 150 GPa (see Figure 5c). Thus, in $[Ac(hex\text{-}H_6)_2](H_2)_6$ Ac atoms along with 24 bounded H-atoms form the hexagonal lattice shown in Figure 5d.

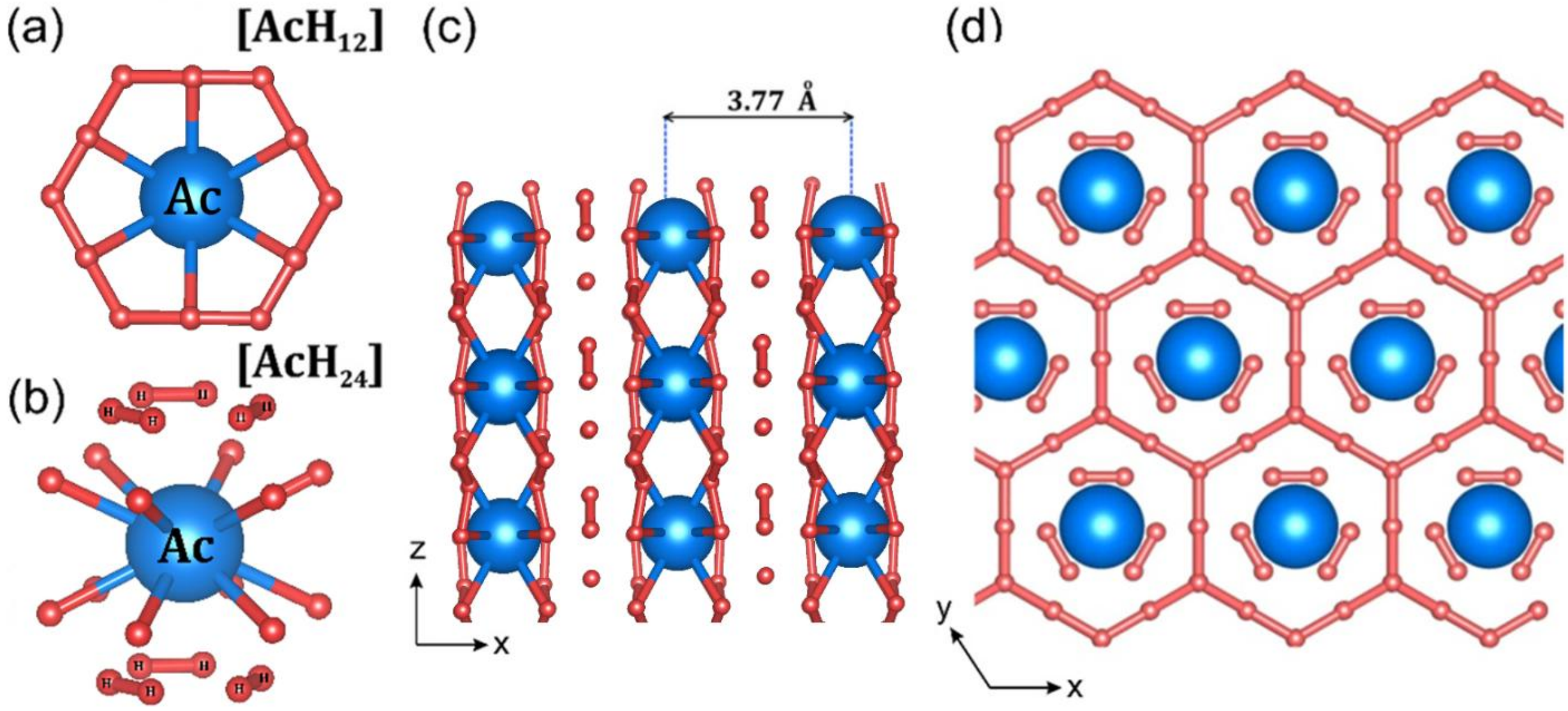

Figure 5. Structural features of $P\bar{6}m2$-$AcH_{16}$ at 150 GPa. a) First coordination sphere of Ac atom with 12 hydrogen atoms; b) Coordination of Ac atom by 24 hydrogen atoms; c) side view of $P\bar{6}m2$-$AcH_{16}$; d) top view of $P\bar{6}m2$-$AcH_{16}$.

Among all predicted hydrogen-rich phases only two are semiconducting (*C2/m*–$AcH_5$ and *Cmc2₁*–$AcH_7$) with DFT band gaps of 0.56 and 1.58 eV, respectively; the systematic underestimation of the band gap by DFT-PBE method should be kept in mind (see Supporting Information for details). The rest of predicted phases are metallic and are potential high-temperature superconductors. Phonon calculations confirmed that the predicted phases have no imaginary phonon frequencies in their predicted ranges of thermodynamic stability (see Supporting Information). The most interesting compounds $AcH_{10}$, $AcH_{12}$ and $AcH_{16}$ demonstrate nonmagnetic behavior. The charge distribution pattern in these compounds shows that H-atoms serve as *e*-donors for Ac and have charge 0.6-0.7$e^-$ while Ac is negatively charged by additional 0.8-1.1 $e^-$ smeared on *d* and *f*-orbitals.

In the framework of our study, the superconducting transition temperature ($T_C$) was calculated using the "full" Allen-Dynes formula (see Eq. (1)) or the Allen-Dynes modified McMillan formula, see Table 2. We calculated the Eliashberg function $\alpha^2F(\omega)$ at different pressures (see Supporting Information). One can note that increase of pressure leads to shifting of $\alpha^2F(\omega)$ function to high frequencies. Our calculations indicate that $\alpha^2F(\omega)$ Eliashberg functions of $R\bar{3}m$-$AcH_{10}$ at 200 GPa (see Figure 6a) and $P\bar{6}m2$-$AcH_{16}$ at 150 GPa (see Supporting Information) have one main "hill". For this type of $\alpha^2F(\omega)$ function the numerical solution of the Eliashberg equations may be approximated by the full Allen-Dynes formula [40]. Calculated $\lambda$ and $\omega_{log}$ as a function of pressure for $R\bar{3}m$-$AcH_{10}$, $I4/mmm$-$AcH_{12}$ and $P\bar{6}m2$-$AcH_{16}$ are summarized in Table 2.

As we mentioned above, the following new stable or conditionally stable phases are metallic and potentially high-$T_C$ superconductors: $Cmcm$-$AcH_3$, $Cmcm$-$AcH_4$, $Cmmm$-$Ac_3H_{10}$, $P\bar{1}$-$AcH_5$, $C2/m$-$AcH_8$, $R\bar{3}m$-$AcH_{10}$, $I4/mmm$-$AcH_{12}$ and $P\bar{6}m2$-$AcH_{16}$. For $AcH_8$, $AcH_{10}$, $AcH_{12}$ and $AcH_{16}$ the EPC coefficient λ exceeds 1.5 and the "full" eq. (1) was used for calculating $T_C$.

$Cmcm$-$AcH_3$ is not a superconductor due to the very low EPC coefficient < 0.1 leading to $T_C$ close to 0 K. Calculations of $T_C$ for $Cmmm$-$Ac_3H_{10}$ and $Cmcm$-$AcH_4$ phase give 3 K (λ = 0.39) and 60 K (λ = 0.89), respectively. This can be explained by increasing of symmetry and hydrogen content versus $Cmcm$-$AcH_3$. However, increase in hydrogen content leads to new phenomena in $AcH_5$: there are two stable phases – semiconducting $C2/m$-$AcH_5$ (stable below 100 GPa) and low-symmetric metallic $P\bar{1}$-$AcH_5$ (above 100 GPa) with relatively high superconducting parameters (λ = 0.92, $\omega_{log}$ = 1162 K). This behavior can be explained by structural differences: in $C2/m$-$AcH_5$ hydrogen sublattice consists of $H_2$ and isolated H-atoms, together forms isolated distorted hexagons while in $P\bar{1}$-$AcH_5$ hydrogen sublattice consists of curved $H_3$ groups which form infinite hydrogen chains.

The same explanation may be applied for $Cmc2_1$-$AcH_7$ (semiconductor, stable < 100 GPa) consisted of isolated $H_2$ and $H_4$-tripods, and $C2/m$-$AcH_8$ (metal, stable > 100 GPa) with unexpectedly high SC parameters (λ = 1.79, $T_C$ = 114-134 K, see Table 2). $AcH_8$ contains curved $H_3$ groups which form well connected branched three-dimensional net.

The highest transition temperature was calculated for $R\bar{3}m$-$AcH_{10}$ to be in a range from 177 to 204 K at 200 GPa (according to Allen-Dynes eq.) depending on the exact $\mu^*$ value (see Table 2 and Figure 6a). Comparison of $\alpha^2F(\omega)$ with the phonon DOS line (light green) shows that low-frequency modes of $H_{32}$-cage provide the main contribution to SC. Effect of high-frequency oscillations of $H_2$-fragments is very weak. Thus, for a certain hydrogen content (Ac:H=1:10), specific for each element, the maximum parameters of the electron-phonon interaction are reached. Further increase of hydrogen content in actinium hydrides ($I4/mmm$-$AcH_{12}$) does not lead to higher $T_C$. Only $P\bar{6}m2$-$AcH_{16}$ below 150 GPa can achieve 200 K (see Figure 6b). As can be seen from Table 2, in the series of hydrides $AcH_{3-10}$ the critical temperature increases monotonically as the concentration of hydrogen atoms increases, but further increase of hydrogen concentration leads to $T_C$ decreasing, which is universal phenomena.

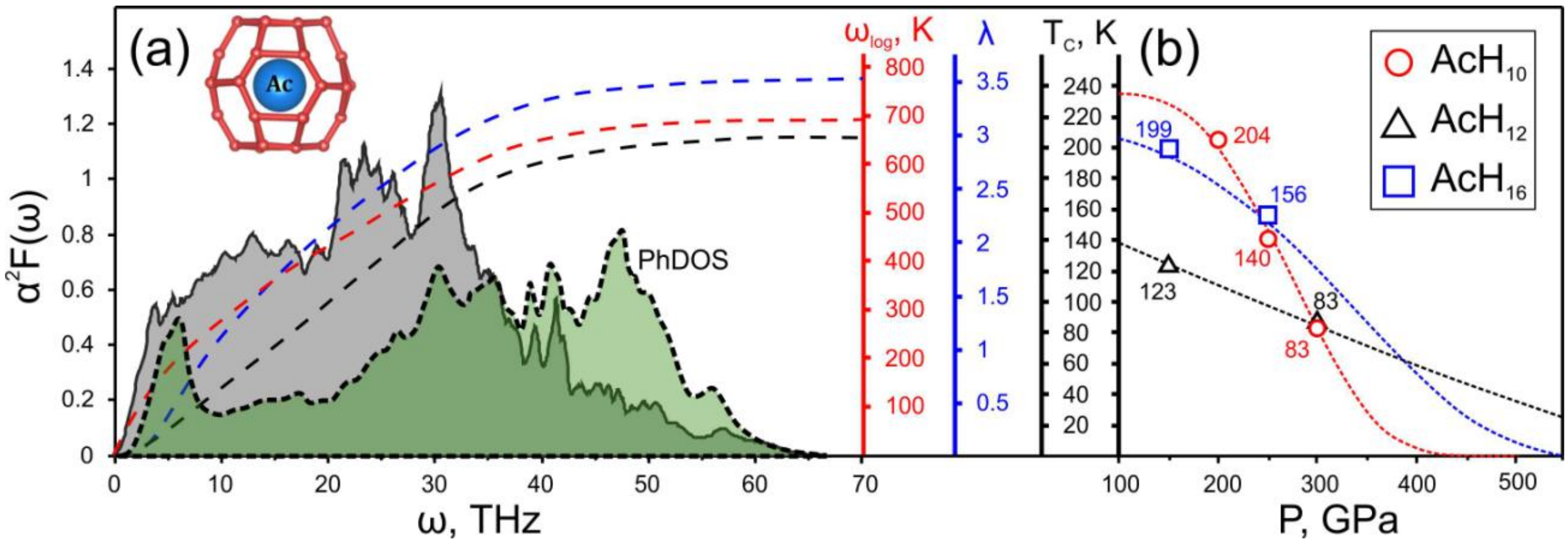

Figure 6. (a) Eliashberg function $\alpha^2F(\omega)$, $\omega_{log}$ , EPC coefficient λ, critical transition temperature ($T_C$) and phonon DOS (a.u.) of $R\bar{3}m$-$AcH_{10}$ at 200 GPa as a function of frequency; (b) $T_C(P)$ functions for $R\bar{3}m$-$AcH_{10}$, $I4/mmm$-$AcH_{12}$ and $P\bar{6}m2$-$AcH_{16}$.

For the hydrogen-richest phases, $R\bar{3}m$-$AcH_{10}$, $I4/mmm$-$AcH_{12}$ and $P\bar{6}m2$-$AcH_{16}$ the dependence of EPC parameters on pressure was studied and extrapolated *via* the well-known empirical equation for low-$T_C$ superconductors [30,31] $-ln\left(\frac{T_C}{\omega_{log}}\right) = Cv^{-\varphi}, C > 0$, where $v$ is the volume (see Ref. [3]), and $C$ and $\varphi$ parameters were calculated earlier for non-transition metals ($\varphi = 2.5 \pm 0.6$)[42,43]. The computed decrease of $T_C$ with pressure for $AcH_{10}$ $dT_C/dP$ = -1.21 K/GPa (at 250 GPa) is much steeper than for $AcH_{12}$ $dT_C/dP$ = -0.263 K/GPa (at 250 GPa), and for $AcH_{16}$ $dT_C/dP$ = -0.433 K/GPa at 250 GPa (see Figure 6b). For comparison, for $CaH_6$ $dT_C/dP$ = -0.33 K/GPa [44]. From calculated data (interpolation of $ln[T_c/\omega_{log}]$ function by $a \times v^b$ law) we can directly determine the coefficients φ =5.53 ($AcH_{10}$), 0.89 ($AcH_{12}$), 1.76 ($AcH_{16}$), which are far from the values for low-temperature superconductors. Obtained extrapolation of $T_C(P)$ is shown in Figure 6b. *I4/mmm*-$AcH_{12}$ was found to be the least sensitive to pressure change. Layer-like character of $P\bar{6}m2$-$AcH_{16}$ leads to substantially higher response to pressure changes in density of states (increases in 2 times in pressure range 150-250 GPa) as well as in EPC parameters (see Table 2). The unusually high derivative $dT_C/dP$ of $AcH_{10}$ may be attributed to metastability of $R\bar{3}m$-$AcH_{10}$ at this pressure (200-250 GPa).

Following the analogy between lanthanum and actinium, in the expected manner, $R\bar{3}m$ - $AcH_{10}$ possesses remarkable high-temperature superconductivity ($T_C$ = 226-251 K) and structural similarity to the recently synthesized $LaH_{10+x}$ [5]. Besides $AcH_{10}$, we found two more superconducting hydrides: *I4/mmm*-$AcH_{12}$ and $P\bar{6}m2$-$AcH_{16}$. Similarity of structures of $LaH_{10}$ and $AcH_{10}$ suggests that La may also have hydrides similar to $AcH_{12}$ and $AcH_{16}$ and these ($LaH_{12}$, $LaH_{16}$) compounds may have been missed by a previous search [1].

**Table 2.** Predicted superconducting properties of actinium hydrides. $T_C$ values are given for $\mu^*$ equal to 0.1, while $T_C$ in brackets are for $\mu^* = 0.15$.

| *Phase* | *P, GPa* | $\lambda$ | $N_f$, *states/atom/eV* | $\omega_{log}$, *K* | $T_C$ *(McMillan), K* | $T_C$ *(Allen-Dynes), K* | $T_C$ [a] *(Eliashberg), K* |
|---|---|---|---|---|---|---|---|
| ***Cmcm*-AcH₃** | 150 | 0.05 | 0.094 | 486 | 0 | 0 | - |
| ***Cmmm*-Ac₃H₁₀** | 150 | 0.39 | 0.071 | 803.6 | 3 (0.55) | 3 (0.55) | - |
| ***Cmcm*-AcH₄** | 100 | 0.89 | 0.071 | 982.5 | 56.5 (41.2) | 60 (43.2) | - |
| $P\bar{1}$**-AcH₅** | 150 | 0.92 | 0.071 | 1162 | 70.5 (49.8) | 74.9 (52) | - |
| ***C2/m*-AcH₈** | 150 | 1.79 | 0.052 | 853 | 113.5 (99.5) | 134 (114.1) | - |
| ***Cm*-AcH₁₀** | 100 | 2.18 | 0.057 | 791.3 | 120.4 (107.9) | 152.1 (130.5) | - |
| $R\bar{3}m$**-AcH₁₀** | 200 | 3.46 | 0.068 | 710.9 | 160.9 (125.5) | 204.1 (177) | 251 (226) |
| | 250 | 1.96 | 0.072 | 812.8 | 115.3 (102.2) | 140.1 (119.7) | 176 (151) |
| | 300 | 1.01 | 0.074 | 1098 | 77.3 (59.7) | 83.2 (63.3) | 101 (76) |
| $I4/mmm$**-AcH₁₂** | 150 | 1.97 | 0.044 | 697 | 99.4 (87.3) | 123.3 (103.7) | 173 (148) |
| | 300 | 1.42 | 0.041 | 677 | 73.4 (62) | 83.8 (68.9) | 101 (81) |
| $P\bar{6}m2$**-AcH₁₆** | 150 | 2.16 | 0.022 | 1054 | 159.7 (143) | 199.2 (171.3) | 241 (221) |
| | 250 | 1.34 | 0.046 | 1383 | 140.5 (118.2) | 155.9 (128.5) | 181 (163) |

[a] Eliashberg equation was solved only for hydrogen-richest phases.

## Conclusions

Exploring the analogy between the electronic structure of actinium and lanthanum atoms, we hypothesized that the Ac-H system may have high-$T_C$ superconducting phases and studied Ac-H system using global optimization within the evolutionary algorithm USPEX at pressures up to 350 GPa. Five new high-temperature superconductors were predicted, namely *C2/m*-$AcH_8$, *Cm*- and $R\bar{3}m$-$AcH_{10}$, $I4/mmm$-$AcH_{12}$ and $P\bar{6}m2$-$AcH_{16}$. $AcH_{10}$ is predicted to be superconducting at temperatures up to 251 K at 200 GPa, $AcH_{16}$ – at temperatures up to 241 K at 110 GPa.

This study is an important part of the work on establishing chemical insight into superconducting properties of binary metal hydrides, which has a clear correlation with the electronic structure of the hydride-forming metal. Namely, we found that the majority of high-$T_C$ superconducting hydrides are concentrated near $d^1f^0$-belt of the Periodic Table among metals with minimal number of *d, f*-electrons. With increasing number of *d, f*-electrons, superconducting properties of metal hydrides become less pronounced, possibly due to increasing Coulomb electron-electron repulsion. Large groups of high-temperature hydride superconductors are tied to specific generative metals lying on the junctions (*p-d, s-p*) of *p*, *d*, *f*-groups of elements.

This work contributes to the establishment of extreme diversity of actinide hydrides, which, as we have shown earlier [2,3], are unexpectedly rich in high-temperature superconducting phases.

## Acknowledgements

The work was supported by Russian Science Foundation (№ 16-13-10459). Calculations were performed on the Rurik supercomputer at MIPT. D.S. expresses his gratitude to the SPbSU resource center "Computer Center SPbU". Authors thank Dr. Piotr M. Kowalski for generation of actinium pseudopotential used in $T_C$ calculations.

## Superconductivity of binary hydrides viewed through the Periodic Table

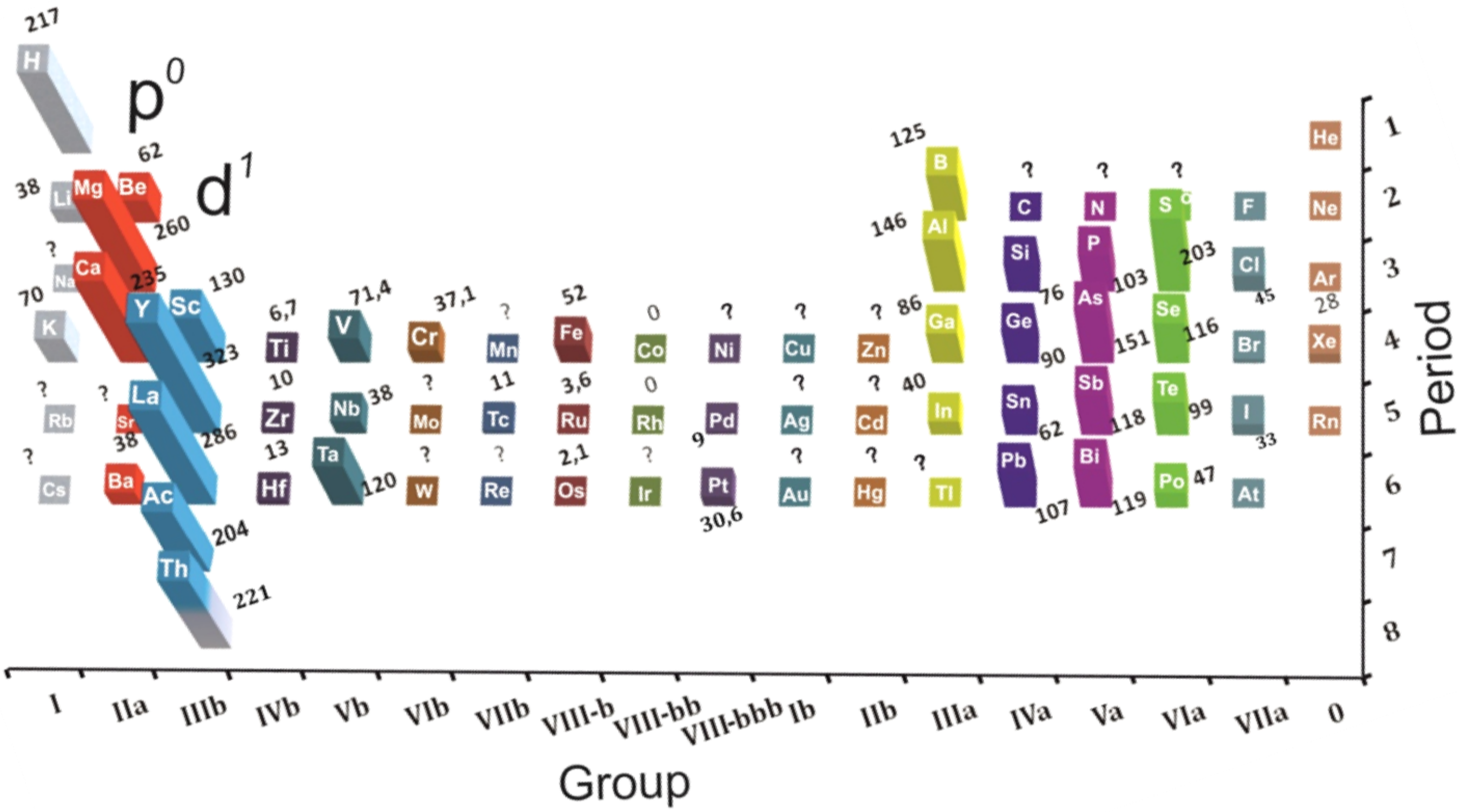

**Figure S1.** Mendeleev table with already investigated binary $XH_n$, where the numbers are maximum calculated Tc. The data were taken from sources given in Table S1.

**Table S1.** Previously predicted or synthesized binary hydrides with maximum $T_C$ values, which were used on Figure S1.

| **Hydride (Pressure, GPa)** | **max $T_c$, K** | **Hydride (Pressure, GPa)** | **max $T_c$, K** | **Hydride (Pressure, GPa)** | **max $T_c$, K** |
|---|---|---|---|---|---|
| $H_{solid}$ (500) | 217 [1] | $TcH_2$ (200) | 11 [14] | $BiH_5$ (300) | 119 [2] |
| $LiH_6$ (150) | 38 [2] | $FeH_5$ (150) | 52 [15] | $H_3S$ (155) | 203 [2] |
| $KH_6$ (166) | 70 [2] | $RuH_3$ (100) | 3.6 [16] | $H_3Se$ (200) | 116 [2] |
| $BeH_2$ (400) | 62 [2] | OsH (100) | 2.1 [17] | $H_4Te$ (200) | 99 [2] |
| $MgH_6$ (300) | 260 [3] | $CoH_3$ | 0 [18] | $H_4Po$ (300) | 47 [23] |
| $CaH_6$ (150) | 235 [2] | RhH | 0 [19] | $H_2Cl$ (450) | 45 [2] |
| $BaH_6$ (100) | 38 [2] | PdH (0) | 9 [20] | $H_2Br$ (240) | 12 [2] |
| $ScH_6$ (285) | 130 [4] | PtH (76) | 30.6 [21] | $H_2I$ (240) | 33 [2] |
| $YH_{10}$ (300) | 323 [5] | $BH_3$ (360) | 125 [2] | XeH (100) | 28 [24] |
| $LaH_{10}$ (300) | 286 [5] | $AlH_5$ (250) | 146 [2] | | |
| $AcH_{10}$ (200) | 204* | $GaH_3$ (160) | 86 [2] | | |
| $ThH_{10}$ (100) | 221 [6] | $InH_3$ (200) | 40 [2] | | |
| $TiH_2$ (0) | 6.7 [7] | $Si_2H_6$ (362) | 79 [22] | | |
| ZrH (120) | 10 [8] | $GeH_8$ (250) | 90 [2] | | |
| $HfH_2$ (260) | 13 [9] | $SnH_4$ (200) | 62 [2] | | |
| $VH_8$ (200) | 71.4 [10] | $PbH_8$ (230) | 107 [2] | | |
| $NbH_4$ (300) | 38 [11] | $PH_3$ (207) | 103 [2] | | |
| $TaH_6$ (50) | 120 [12] | $AsH_8$ (350) | 151 [2] | | |
| $CrH_3$ (81) | 37.1 [13] | $SbH_4$ (150) | 118 [2] | | |

*This work, Allen-Dynes equation.

## Thermodynamic parameters of Ac-H phases

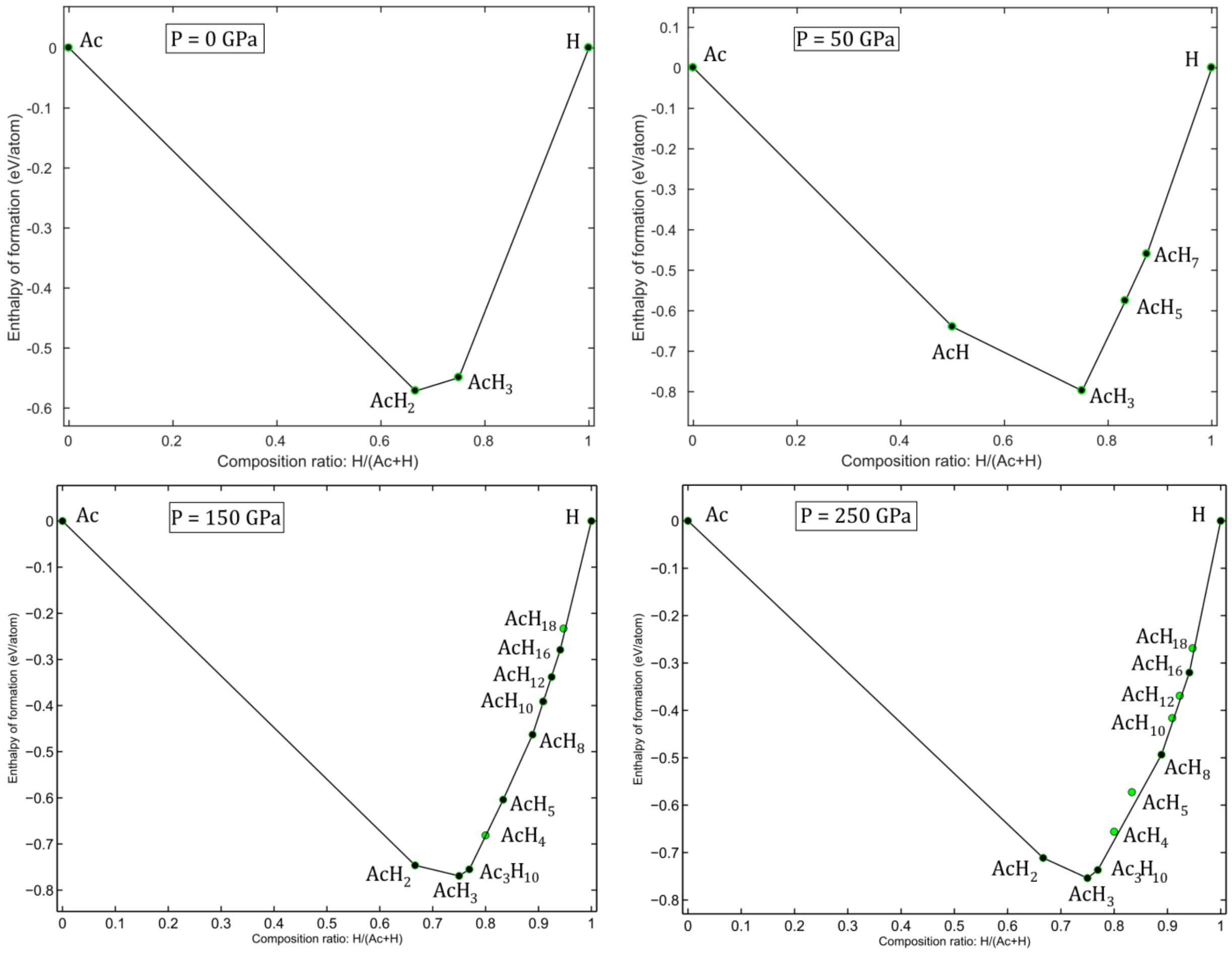

**Figure S2.** Convex Hulls of Ac-H system for 0, 50, 150, 250 GPa.

**Table S3.** Zero-point energies and thermal part of the Helmholtz free energy ($F_{th}$)* of the most interesting Ac-H phases at 150 GPa.

| | **Pressure: 150 GPa** | | | | |
|---|---|---|---|---|---|
| **Temperature** | **0 K** | **300 K** | **500 K** | **1000 K** | **1500 K** |
| | **ZPE, eV/atom** | | **$F_{th}$, eV/atom** | | |
| $Cmcm$-**$AcH_3$** | 0.1358 | -0.0297 | -0.0738 | -0.2248 | -0.4107 |
| $Cmmm$-**$Ac_3H_{10}$** | 0.1375 | -0.0265 | -0.0681 | -0.2124 | -0.3916 |
| $P\bar{1}$-**$AcH_5$** | 0.1388 | -0.0223 | -0.0608 | -0.1972 | -0.3684 |
| $C2/m$-**$AcH_8$** | 0.1394 | -0.0333 | -0.0778 | -0.2264 | -0.4072 |
| $R\bar{3}m$-**$AcH_{10}$** | 0.1405 | -0.0329 | -0.0781 | -0.2300 | -0.4154 |
| $I4/mmm$-**$AcH_{12}$** | 0.1519 | -0.0330 | -0.0802 | -0.2397 | -0.4349 |
| $P\bar{6}m2$-**$AcH_{16}$** | 0.159 | -0.0297 | -0.0738 | -0.2248 | -0.4107 |

$$* \; F_{th} = k_B T \cdot \int_{\Omega} PHDOS(\omega) \cdot \ln\left[1 - \exp\left(-\frac{\hbar\omega}{k_B T}\right)\right] d\hbar\omega$$

**Table S4.** Zero-point energies and thermal part of the Helmholtz free energy of the main superconducting Ac-H phases at 250 GPa.

| | Pressure: 250 GPa | | | | |
|---|---|---|---|---|---|
| **Ttemperature** | **0 K** | **300 K** | **500 K** | **1000 K** | **1500 K** |
| | **ZPE, eV/atom** | **$F_{th}$, eV/atom** | | | |
| $R\bar{3}m$-$AcH_{10}$ | 0.1624 | -0.0248 | -0.0643 | -0.2035 | -0.3776 |
| $I4/mmm$-$AcH_{12}$ | 0.1763 | -0.0238 | -0.0620 | -0.1977 | -0.3682 |
| $P\bar{6}m2$-$AcH_{16}$ | 0.1821 | -0.0207 | -0.0563 | -0.1854 | -0.3493 |

**Table S5.** Calculated enthalpies of Ac-H phases (per formula unit) associated with $AcH_2$ transformations at T = 0 K. H( $Fm\bar{3}m$ -Ac) = -4.092 eV/atom.

| | H, eV | | | | | |
|---|---|---|---|---|---|---|
| **P, GPa** | $Fm\bar{3}m$ - $AcH_2$ | $P4_2/mmc$- $AcH_2$ | $Imma$-$AcH_2$ | $P6_3/mmm$ -$AcH_2$ | $Fm\bar{3}m$ - $AcH_3$ | $Fm\bar{3}m$ - AcH |
| 0 | -12.591 | -12.242 | -12.188 | -11.107 | -16.443 | -16.443 |
| 10 | -9.599 | -9.531 | -9.468 | -8.806 | -13.565 | -5.711 |
| 20 | -6.934 | -7.103 | -7.072 | -6.696 | -10.952 | -3.568 |
| 30 | -4.491 | -4.871 | -4.889 | -4.601 | -8.521 | -1.600 |
| 40 | -2.234 | -2.786 | -2.863 | -2.672 | -6.232 | 0.212 |
| 50 | -0.113 | -0.816 | -0.953 | -0.821 | -4.055 | 1.917 |
| 60 | 1.893 | 1.058 | 0.857 | 0.903 | -1.973 | 3.534 |
| 70 | 3.789 | 2.855 | 2.588 | 2.544 | 0.0285 | 5.078 |

**Table S6.** Calculated enthalpies of Ac-H phases (per formula unit) associated with $AcH_4$ decomposition at T = 0 K.

| | H, eV | | | |
|---|---|---|---|---|
| **P, GPa** | $Cmcm$-$AcH_4$ | $Cmmm$-$Ac_3H_{10}$ | $P\bar{1}$-$AcH_5$ | $C2/m$-$AcH_8$ |
| 100 | 3.529 | 13.413 | 2.109 | -1.777 |
| 120 | 6.926 | 23.034 | 5.776 | 2.602 |
| 140 | 10.161 | 32.205 | 9.273 | 6.773 |
| 160 | 13.264 | 41.034 | 12.615 | 10.760 |
| 180 | 16.235 | 49.505 | 15.850 | 14.589 |
| 200 | 19.125 | 57.692 | 18.975 | 18.285 |
| 220 | 21.927 | 65.628 | 22.006 | 21.866 |
| 240 | 24.658 | 73.342 | 24.949 | 25.347 |

**Table S7.** Calculated enthalpies of Ac-H phases (per formula unit) associated with $AcH_{12}$ decomposition at T = 0 K.

| | H, eV | | | |
|---|---|---|---|---|
| **P, GPa** | $R\bar{3}m$ **-AcH$_{10}$** | $C2/m$ **-AcH$_8$** | $P\bar{6}m2$**-AcH$_{16}$** | $I4/mmm$**-AcH$_{12}$** |
| 150 | 7.56 | 8.789 | 3.85 | 6.383 |
| 170 | 11.93 | 12.693 | 9.551 | 11.203 |
| 190 | 16.141 | 16.453 | 15.044 | 15.836 |
| 210 | 20.219 | 20.089 | 20.354 | 20.328 |
| 230 | 24.178 | 23.618 | 25.5 | 24.684 |
| 250 | 28.028 | 27.053 | 30.5 | 28.919 |
| 270 | 31.78 | 30.400 | 35.367 | 33.046 |
| 290 | 35.441 | 33.665 | 40.114 | 37.072 |
| 310 | 39.07 | 36.857 | 44.753 | 41.006 |
| 330 | 42.558 | 39.983 | 49.292 | 44.856 |
| 350 | 45.973 | 43.046 | 53.738 | 48.636 |

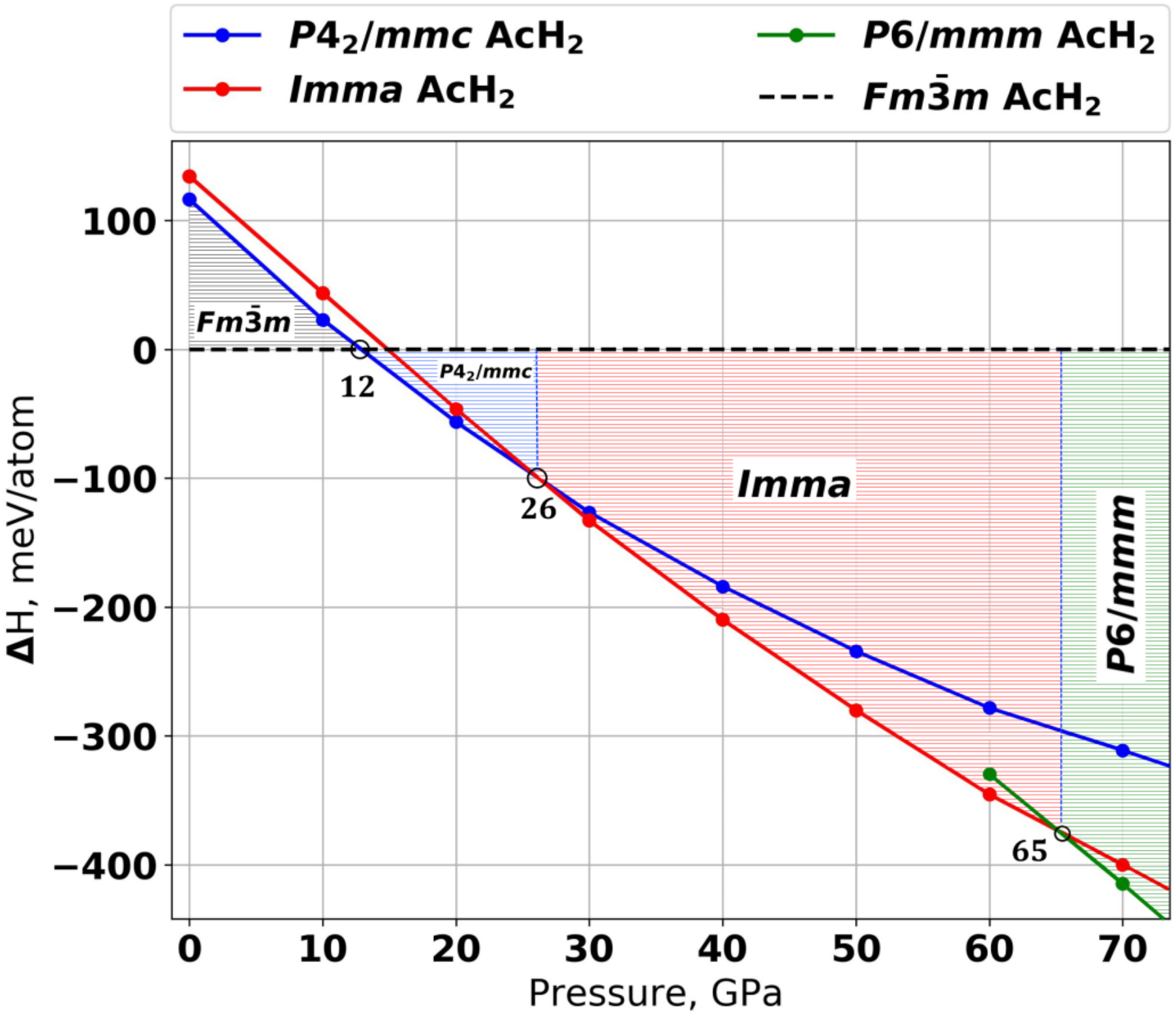

**Figure S3.** Enthalpy diagram for different crystal structures of $AcH_2$.

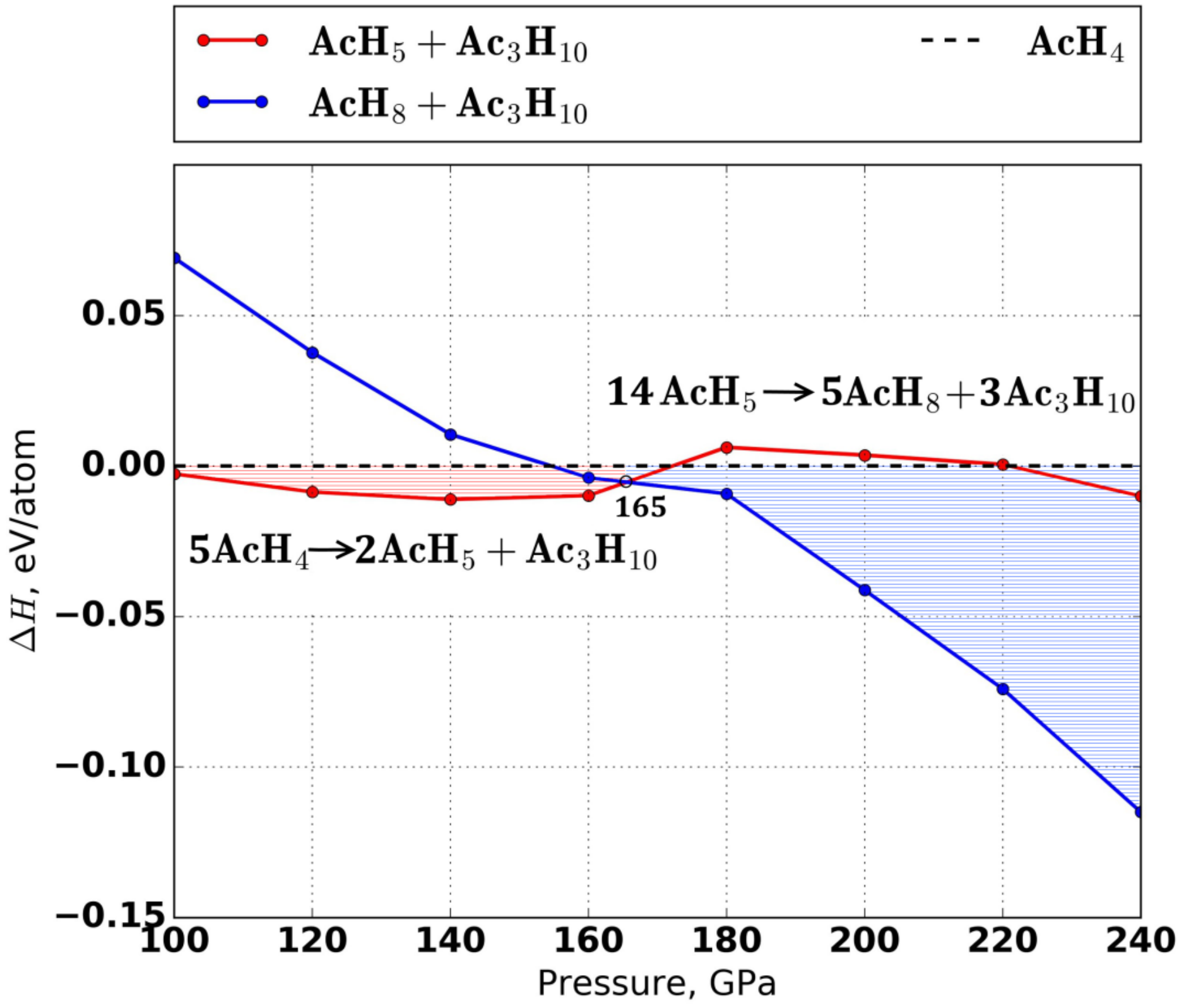

**Figure S4.** Enthalpy diagram for chemical reactions of $AcH_4$ and $AcH_5$ under pressure.

## Equations for calculating $T_C$

At the first stage of self-consistent procedure of the numerical solution of the Eliashberg equation [25] we calculated Fredholm kernels of electron-phonon interaction $I_1(\omega,\omega')$, $L_{-1}(\omega,\omega')$, $I_2(\omega,\omega')$. Integration was divided into intervals using the following symmetry properties $\Delta(-\omega) = \Delta^*(\omega)$, $Z(-\omega) = Z^*(\omega)$, $\alpha^2(-\omega)F(-\omega) = \alpha^2(\omega)F(\omega)$ [26], but avoiding points T = 0 K, ω = 0 Ry due to possible divergences of 1/T, 1/ω, ω/T during numerical integration [26-27]:

$$I_1(\omega,\omega') = \int_0^{\infty} \alpha^2(\Omega)F(\Omega)\cdot\left\{\frac{\left(\exp\left(-\frac{\hbar\omega'}{kT}\right)+1\right)^{-1}+\left(\exp\left(\frac{\hbar\Omega}{kT}\right)-1\right)^{-1}}{\hbar\omega'+\hbar\Omega-\hbar\omega}+\frac{\left(\exp\left(\frac{\hbar\omega'}{kT}\right)+1\right)^{-1}+\left(\exp\left(\frac{\hbar\Omega}{kT}\right)-1\right)^{-1}}{\hbar\omega'-\hbar\Omega-\hbar\omega}\right\}\cdot\hbar d\Omega$$

$$I_{-1}(\omega,\omega') = \int_0^{\infty} \alpha^2(\Omega)F(\Omega)\cdot\left\{\frac{\left(\exp\left(\frac{\hbar\omega'}{kT}\right)+1\right)^{-1}+\left(\exp\left(\frac{\hbar\Omega}{kT}\right)-1\right)^{-1}}{-\hbar\omega'+\hbar\Omega-\hbar\omega}-\frac{\left(\exp\left(-\frac{\hbar\omega'}{kT}\right)+1\right)^{-1}+\left(\exp\left(\frac{\hbar\Omega}{kT}\right)-1\right)^{-1}}{\hbar\omega'+\hbar\Omega+\hbar\omega}\right\}\cdot\hbar d\Omega$$

$$\frac{I_2(\omega,\omega')}{\hbar\omega} = \int_0^{\infty} 2\alpha^2(\Omega)F(\Omega)\cdot\left\{\frac{\left(\exp\left(\frac{\hbar\omega'}{kT}\right)+1\right)^{-1}+\left(\exp\left(\frac{\hbar\Omega}{kT}\right)-1\right)^{-1}}{(\hbar\omega)^2-(\hbar\Omega-\hbar\omega')^2}-\frac{\left(\exp\left(-\frac{\hbar\omega'}{kT}\right)+1\right)^{-1}+\left(\exp\left(\frac{\hbar\Omega}{kT}\right)-1\right)^{-1}}{(\hbar\Omega+\hbar\omega')^2-(\hbar\omega)^2}+\right\}\cdot\hbar d\Omega$$

Then, we calculated the Coulomb contribution to the order parameter using common expression for electronic density of states in BCS theory [26]:

$$I_3 = -\mu\cdot\int_0^{\infty}\mathrm{Re}\left(\frac{\Delta(\omega')}{\sqrt{\hbar^2\omega'^2-\Delta^2(\omega')}}\right)\cdot\frac{\exp\left(\frac{\hbar\omega'}{kT}\right)-1}{\exp\left(\frac{\hbar\omega'}{kT}\right)+1}\cdot\hbar d\omega'\cdot$$

Calculation of average Coulomb electron-electron repulsion was performed using the upper bound of empirical value $\mu^* = 0.1$ (0.15) with $\omega_c = 75$ THz = 0.31 eV (upper bound of phonon spectrum) and $E_e = 85$ eV (characteristic width of electron spectrum in $AcH_{10}$) by a common way gives $\mu = 0.23$ (0.95).

$$\mu = \frac{\mu^*}{1-\mu^*\ln\frac{E_e}{\hbar\omega_c}}$$

Then, a function of electron mass renormalization $Z(\omega)$ was calculated as

$$Z(\omega) = 1-\int_0^{\infty}\mathrm{Re}\left(\frac{\hbar\omega'}{\sqrt{\hbar^2\omega'^2-\Delta^2(\omega')}}\right)\cdot\frac{I_2(\omega,\omega')}{\hbar\omega}\cdot\hbar d\omega',$$

After that the next approximation for the order parameter (or the value of superconducting gap, Δ(ω))

$$\Delta(\omega)=\frac{I_3+\int_0^{\infty}\mathrm{Re}\left(\frac{\Delta(\omega')}{\sqrt{\hbar^2\omega'^2-\Delta^2(\omega')}}\right)\cdot I_1(\omega,\omega')\,\hbar d\omega'+\int_0^{\infty}\mathrm{Re}\left(\frac{\Delta(\omega')}{\sqrt{\hbar^2\omega'^2-\Delta^2(\omega')}}\right)\cdot I_{-1}(\omega,\omega')\,\hbar d\omega'}{Z(\omega)}$$

and the new $I_1(\omega,\omega')$, $I_{-1}(\omega,\omega')$, $I_2(\omega,\omega')$ integrals are calculated.

Then, the next iteration for Δ(ω) is calculated and the new cycle starts again. Value of Δ(ω) obtained after 10-20 iterations was used for construction of $\Delta(T, \omega)|_{\omega\to0} = \Delta(T)$ function and search for Tc (Δ(Tc) →0) as well as characteristic ratio 2Δ(0)/Tc and its deviation from weak coupling limit (3.52).

**Table S8.** Comparison of critical temperatures calculated *via* different methods for $R\bar{3}m$-$AcH_{10}$, *I4/mmm*-$AcH_{12}$, $P\bar{6}m2$-$AcH_{16}$. $T_C$ values are given for $\mu^*$ in the range of 0.1-0.15, Eliashberg equation was solved for $\mu^* = 0.1$ ($\mu = 0.23$) and 0.15 ($\mu = 0.95$).

| Phase | *P*, GPa | λ | $\omega_{log}$, K | $T_C$ (Allen-Dynes), K | $T_C$ (McMillan), K | Tc (Eliashberg),* K |
|---|---|---|---|---|---|---|
| $AcH_{10}$ | 200 | 3.46 | 710.9 | 204.1 (177) | 160.9 (125.5) | 251 (226) |
| | 250 | 1.96 | 812.8 | 140.1 (119.7) | 115.3 (102.2) | 176 (151) |
| | 300 | 1.01 | 1098.0 | 83.2 (63.3) | 77.3 (59.7) | 101 (76) |
| $AcH_{12}$ | 150 | 1.97 | 697.0 | 123.3 (103.7) | 99.4 (87.3) | 173 (148) |
| | 300 | 1.42 | 677 | 83.8 (68.9) | 73.4 (62) | 101 (81) |
| $AcH_{16}$ | 150 | 2.16 | 1054 | 199.2 (171.3) | 159.7 (143) | 241 (221) |
| | 250 | 1.34 | 1382.7 | 155.9 (128.5) | 140.5 (118.2) | 181 (163) |

## Phonon spectra of Ac-H phases

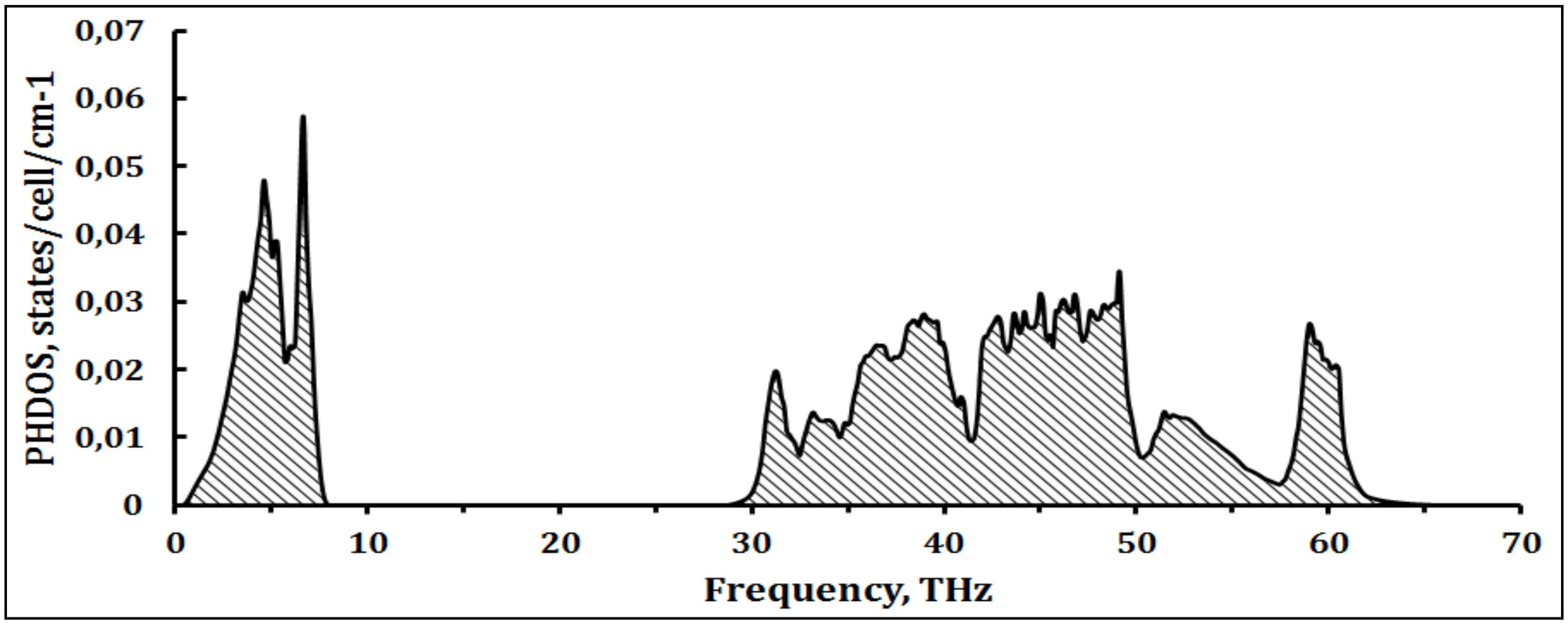

**Figure S19.** Phonon density of states of $Cmcm$-$AcH_3$ at 150 GPa.

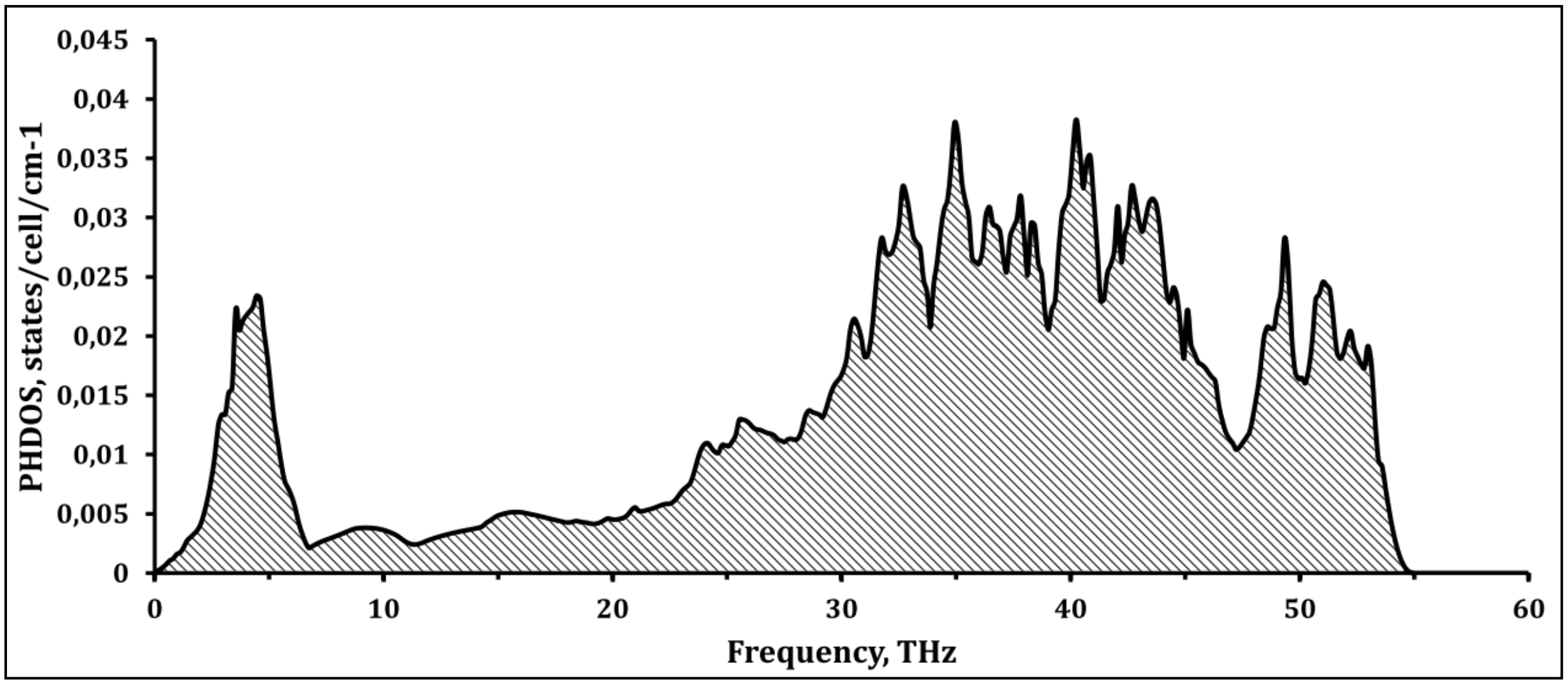

**Figure S20.** Phonon density of states of $Cmcm$-$AcH_4$ at 100 GPa.

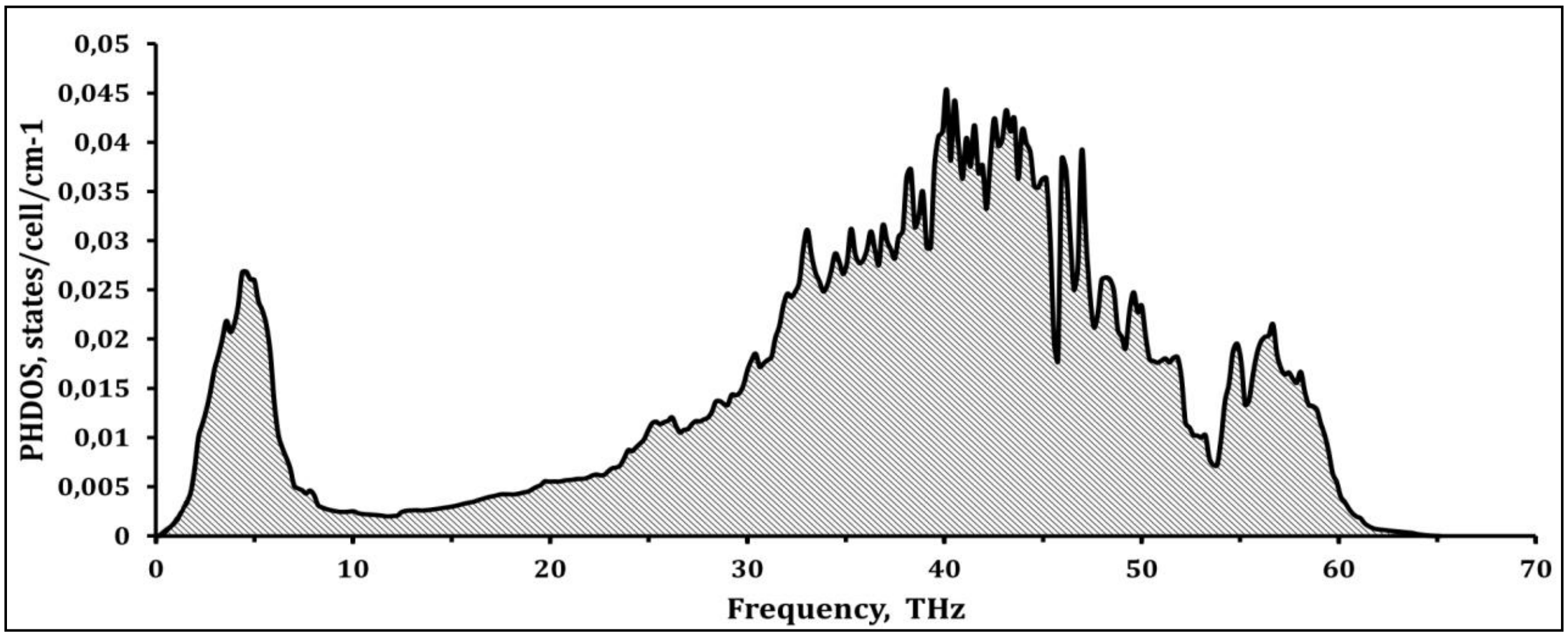

**Figure S21.** Phonon density of states of $P\bar{1}$-$AcH_5$ at 150 GPa.

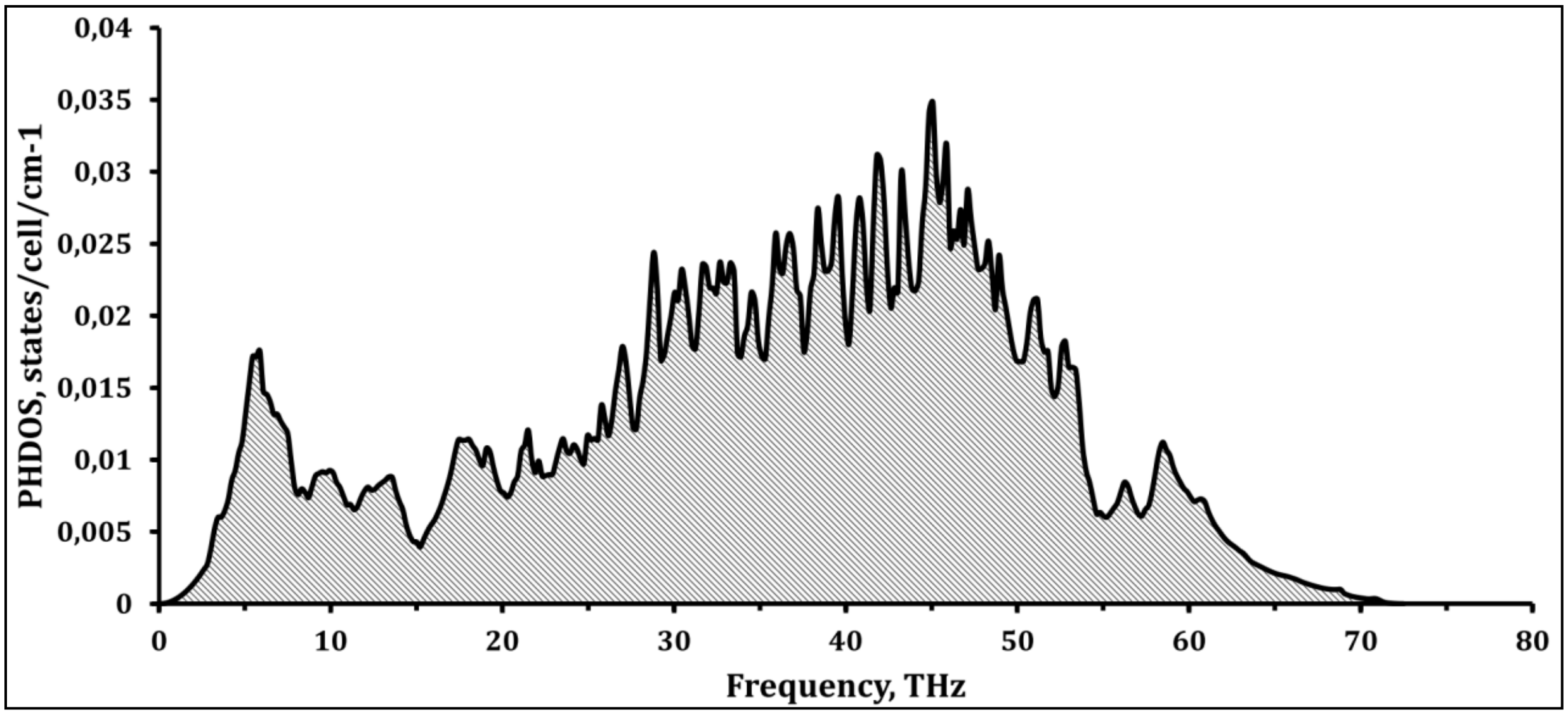

**Figure S22.** Phonon density of states of $Cm$-$AcH_{10}$ at 100 GPa.

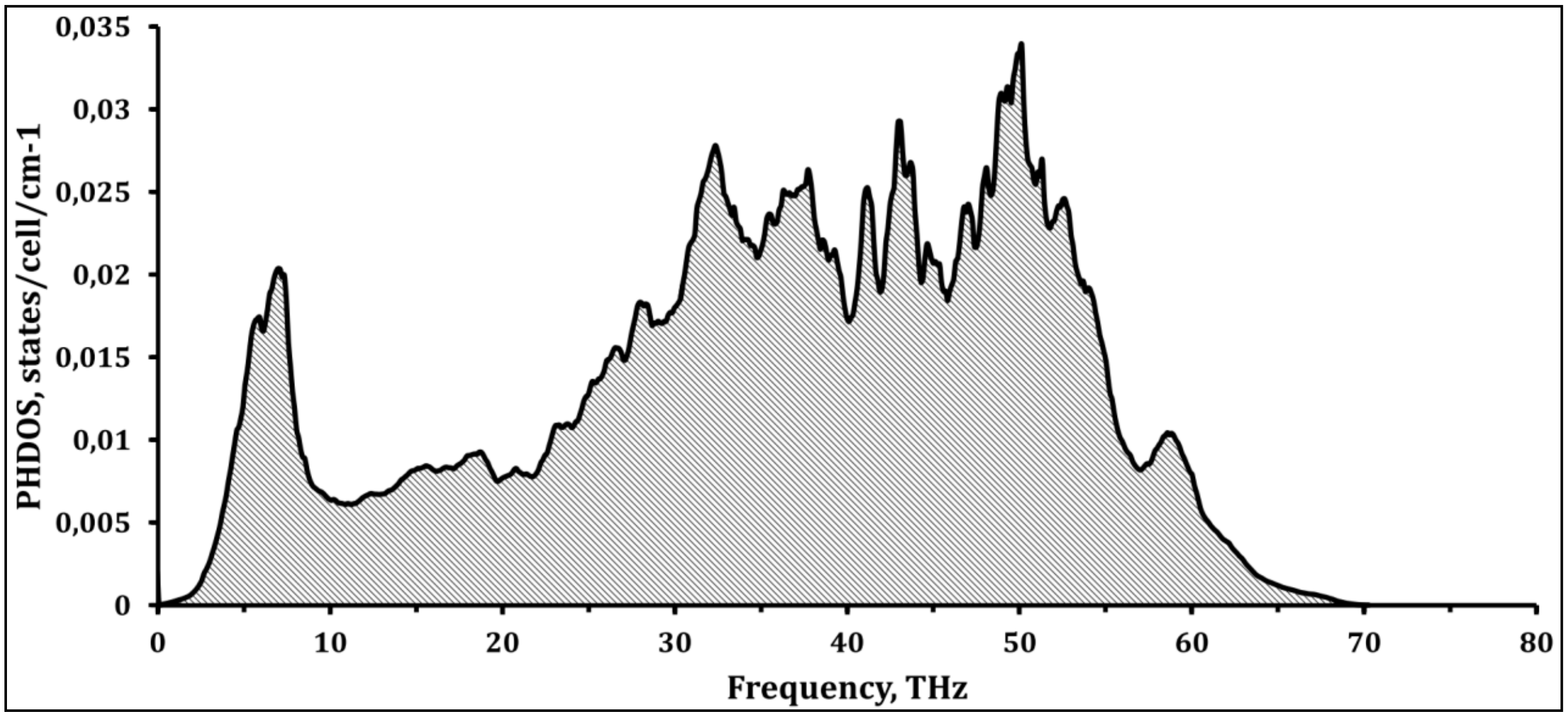

**Figure S23.** Phonon density of states of $R\overline{3}m$-$AcH_{10}$ at 200 GPa.

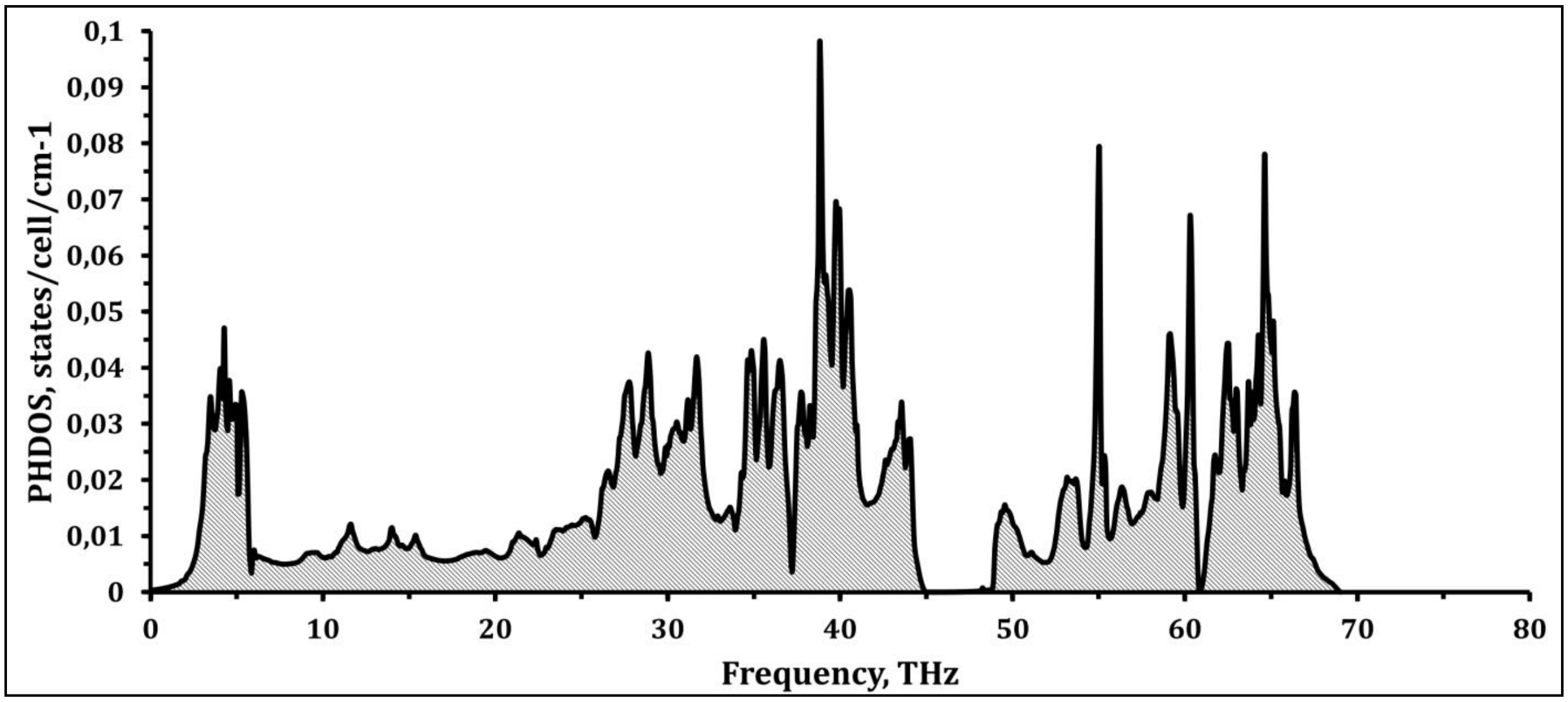

**Figure S24.** Phonon density of states of $I4/mmm$-$AcH_{12}$ at 150 GPa.

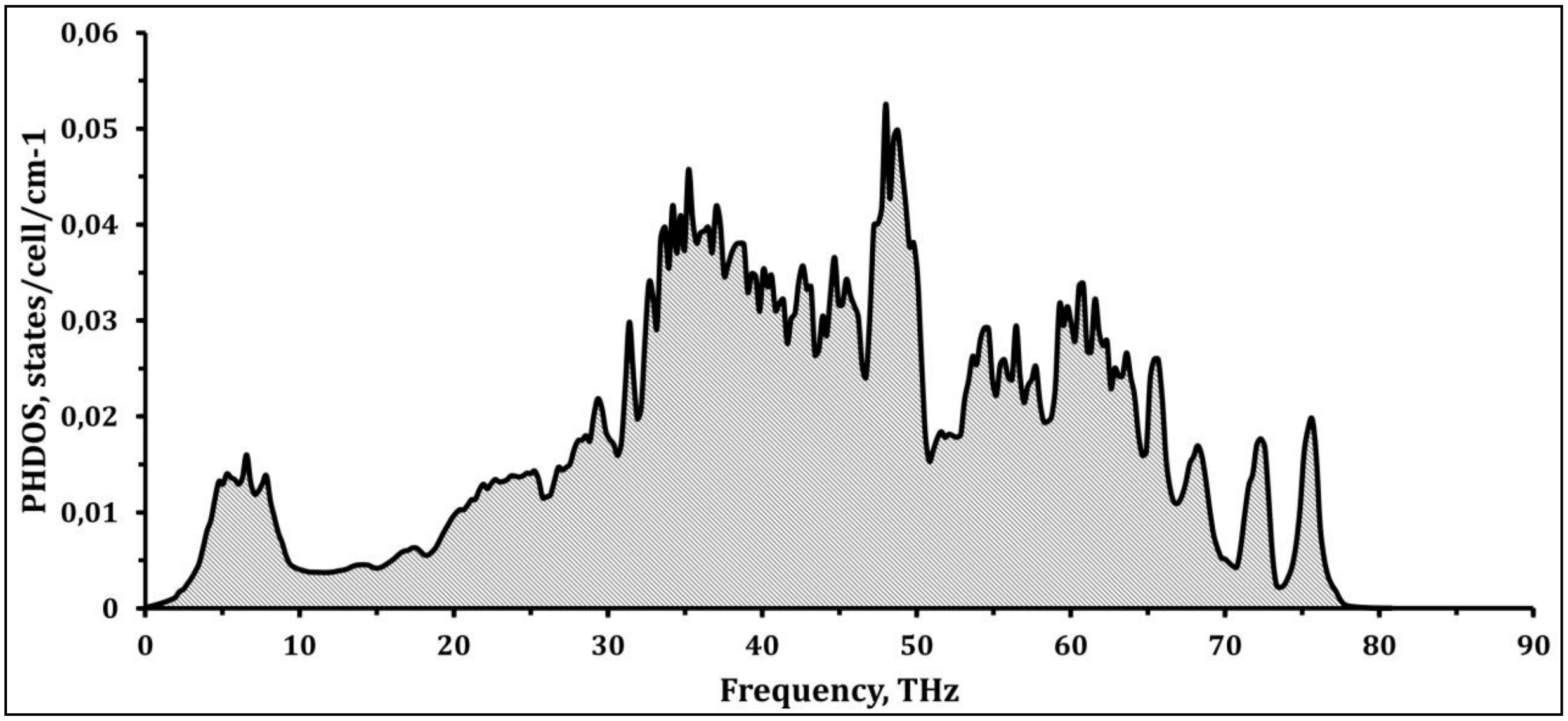

**Figure S25.** Phonon density of states of $P\bar{6}m2$-$AcH_{16}$ at 250 GPa.

# Eliashberg spectral functions for metallic Ac-H phases

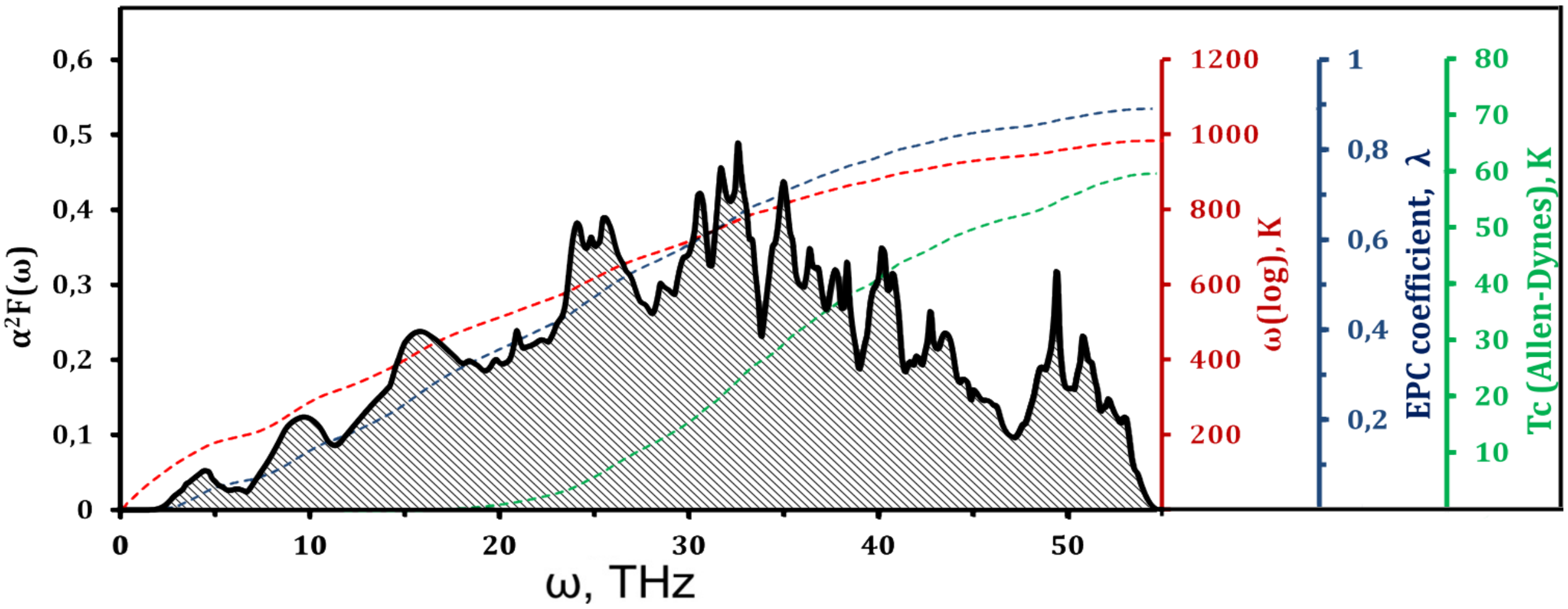

**Figure S26.** Eliashberg function $a^2F(\omega)$ of $Cmcm$-$AcH_4$ at 100 GPa.

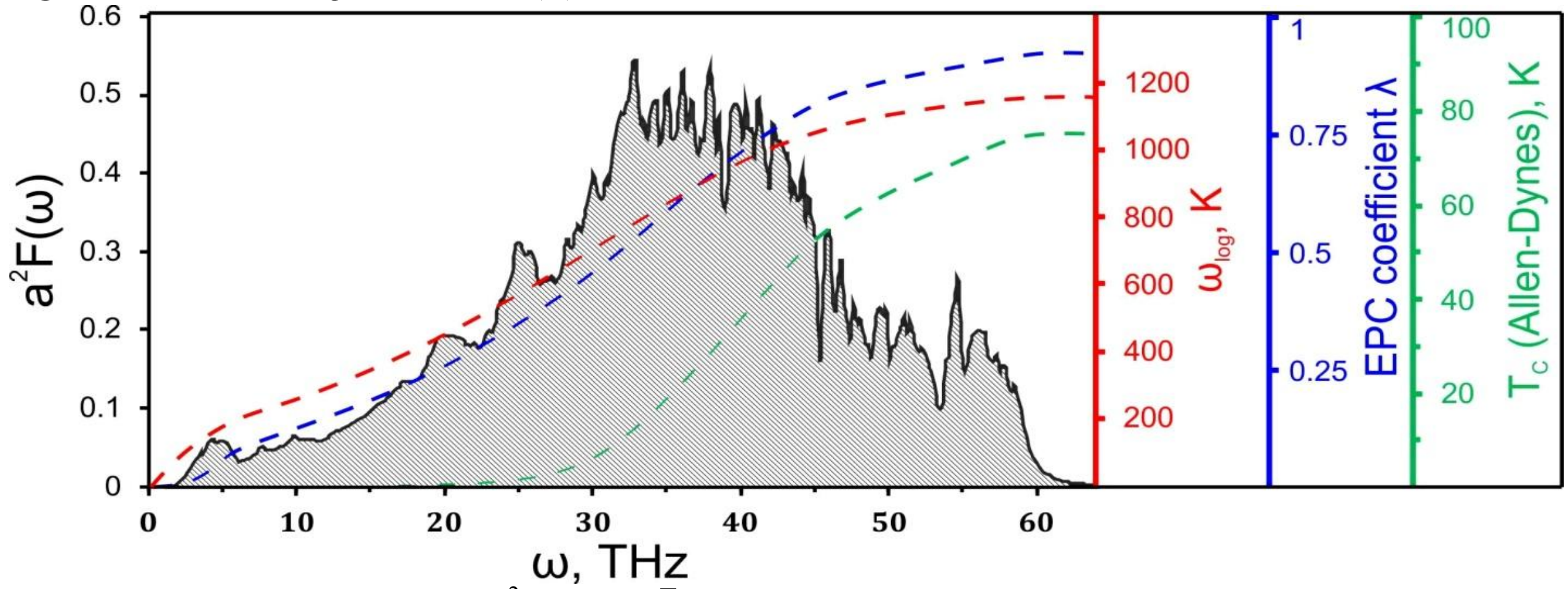

**Figure S27.** Eliashberg function $a^2F(\omega)$ of $P\bar{1}$-$AcH_5$ at 150 GPa.

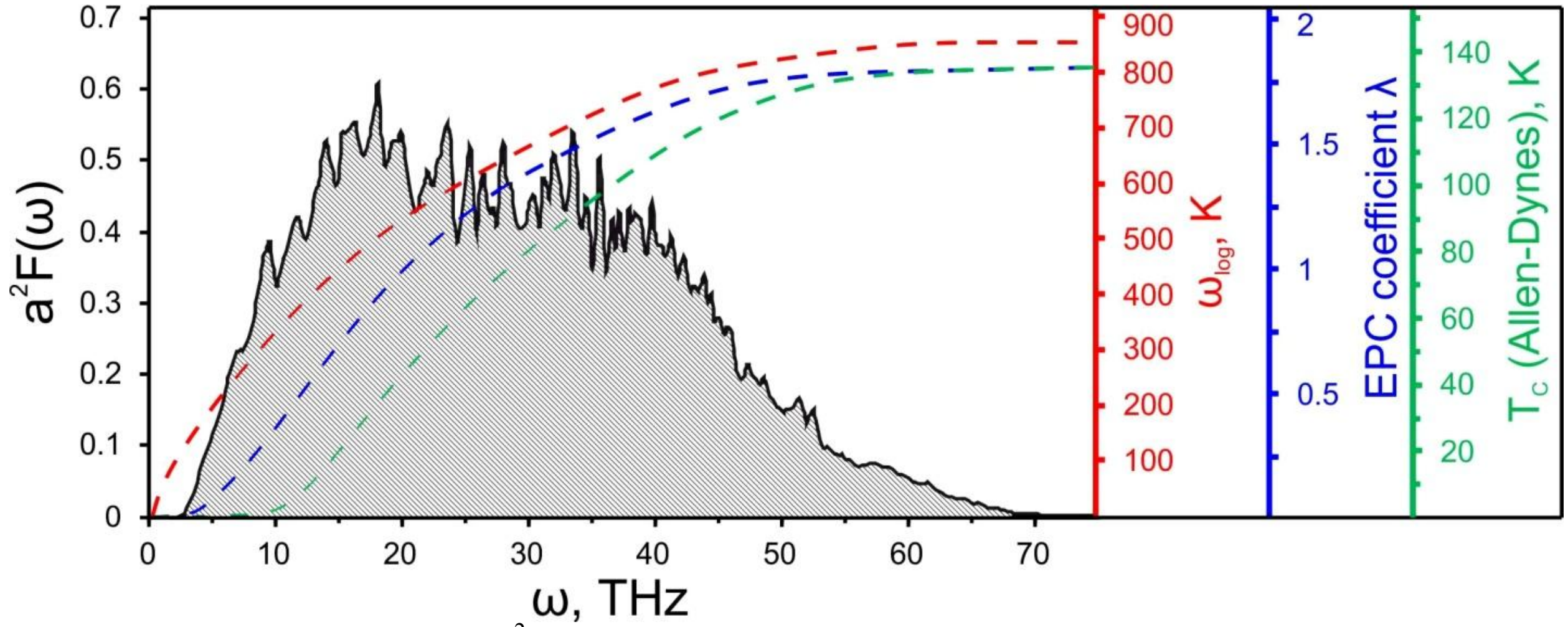

**Figure S28.** Eliashberg function $a^2F(\omega)$ of $C2/m$-$AcH_8$ at 150 GPa.

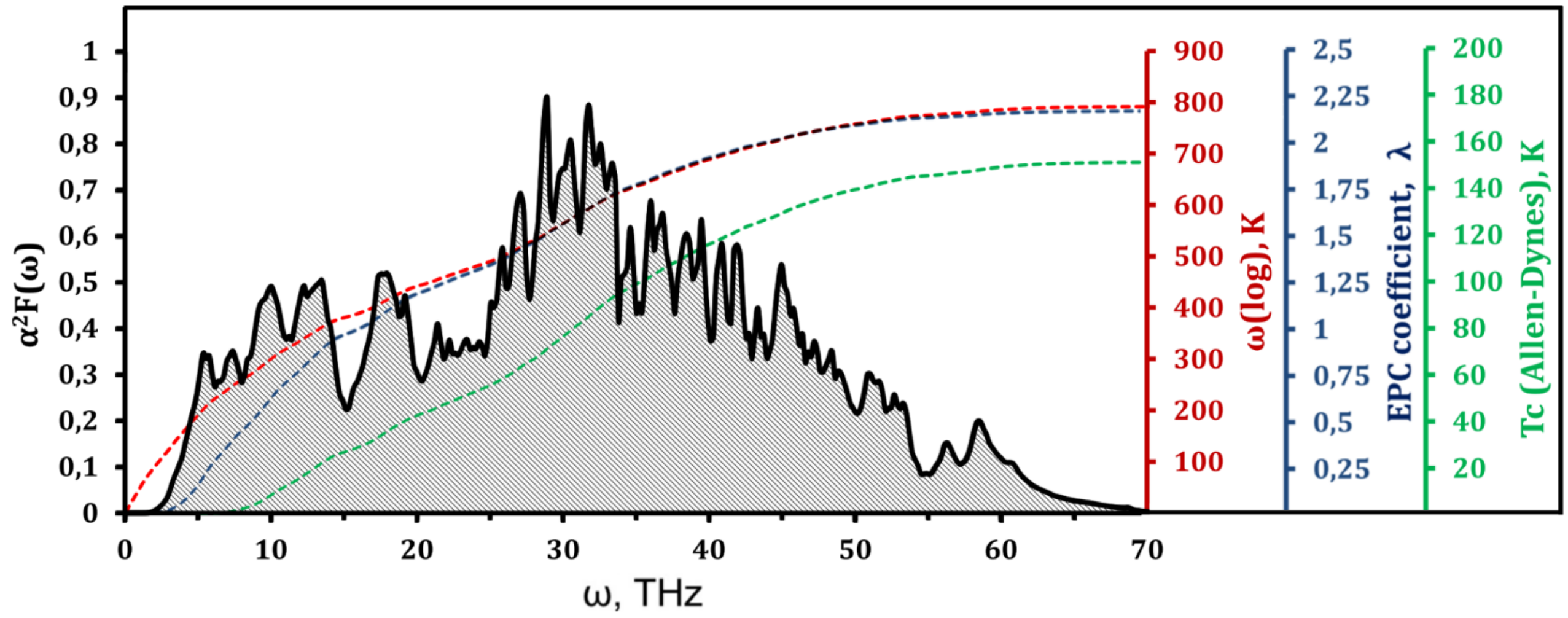

**Figure S29.** Eliashberg function $a^2F(\omega)$ of $Cm$-$AcH_{10}$ at 100 GPa.

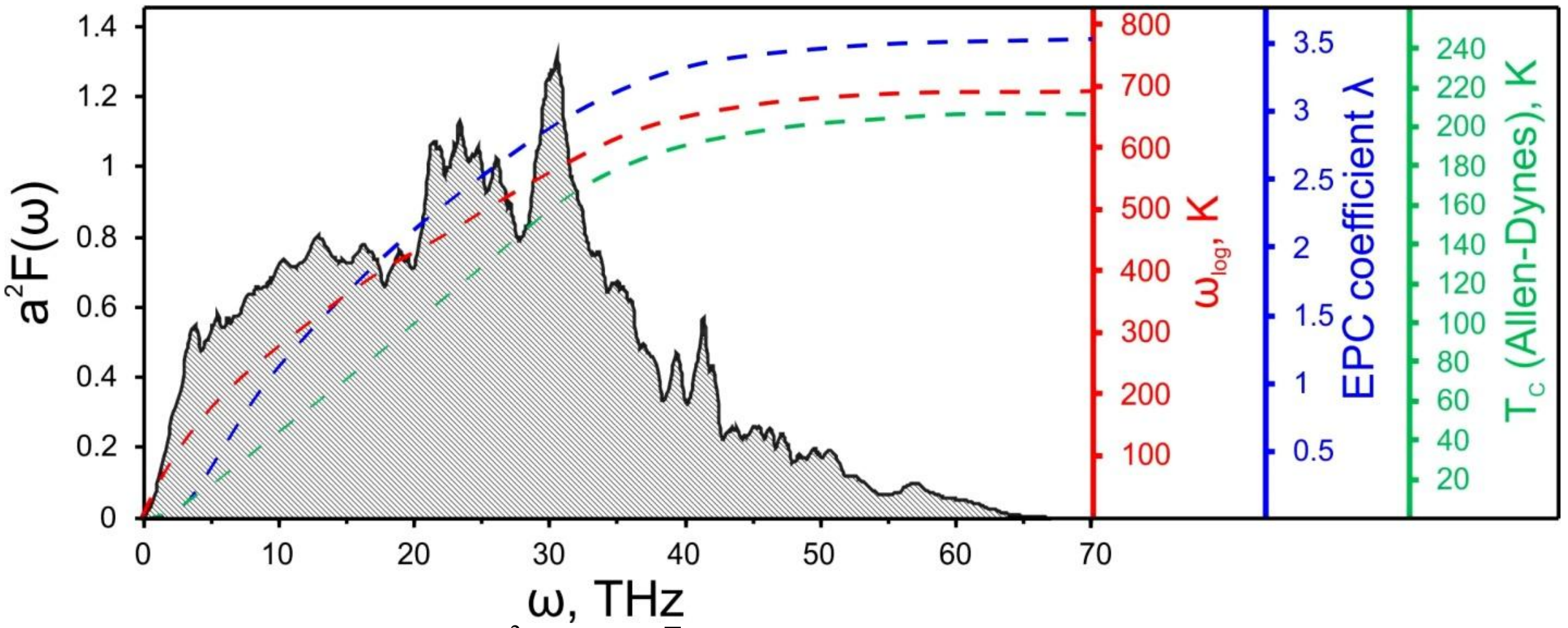

**Figure S30.** Eliashberg function $a^2F(\omega)$ of $R\overline{3}m$ -$AcH_{10}$ at 200 GPa.

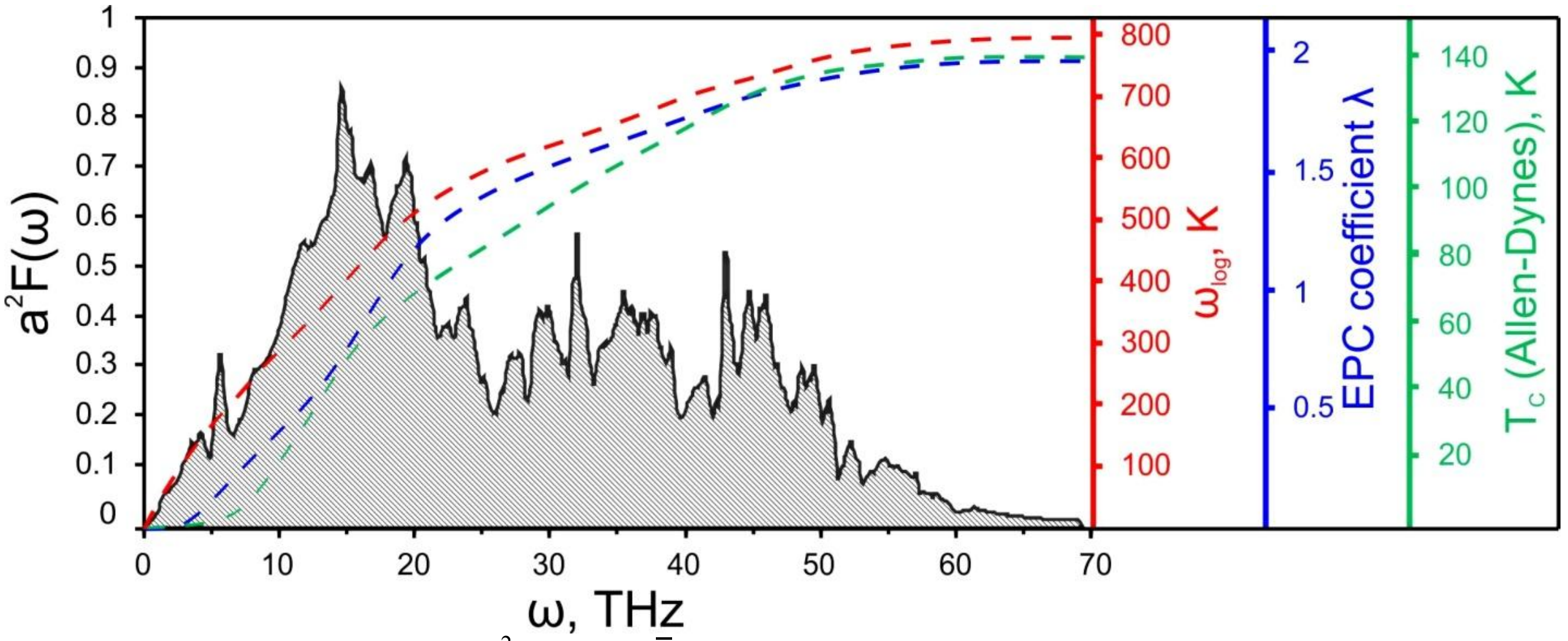

**Figure S31.** Eliashberg function $a^2F(\omega)$ of $R\overline{3}m$-$AcH_{10}$ at 250 GPa.

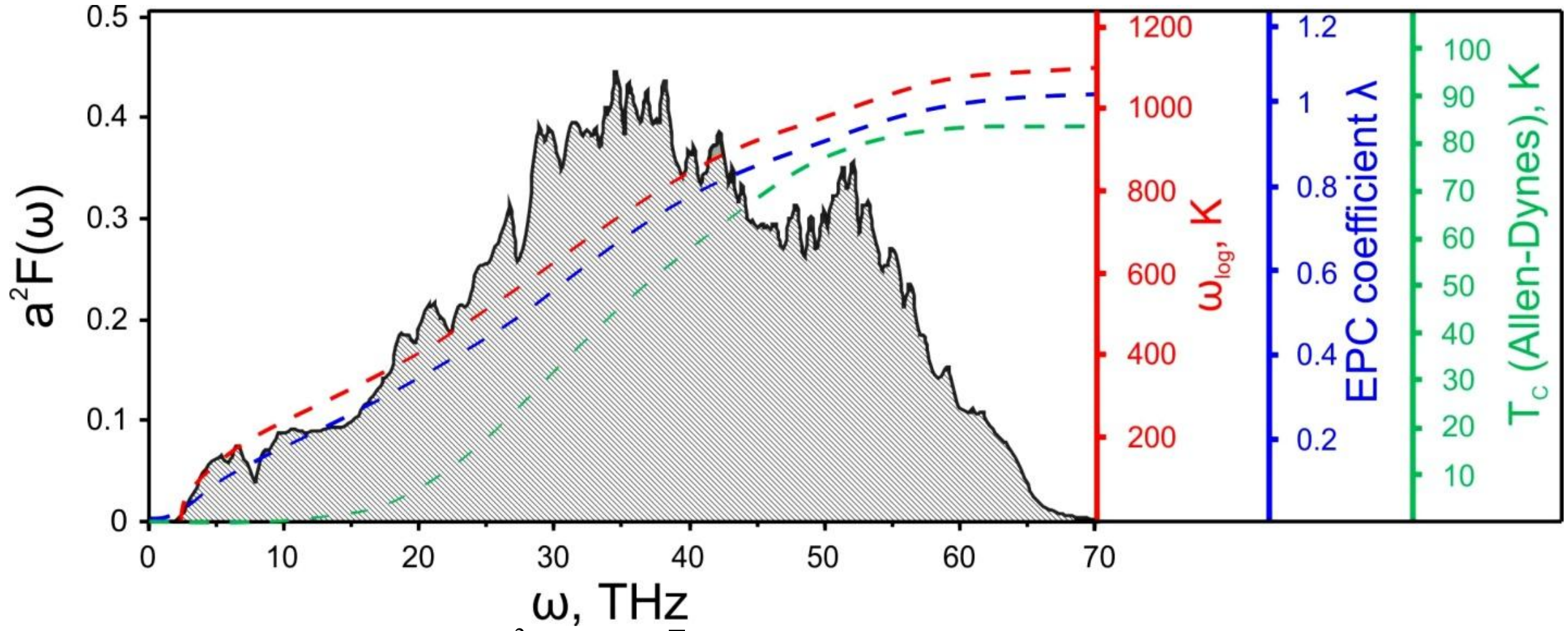

**Figure S32.** Eliashberg function $a^2F(\omega)$ of $R\overline{3}m$-$AcH_{10}$ at 300 GPa.

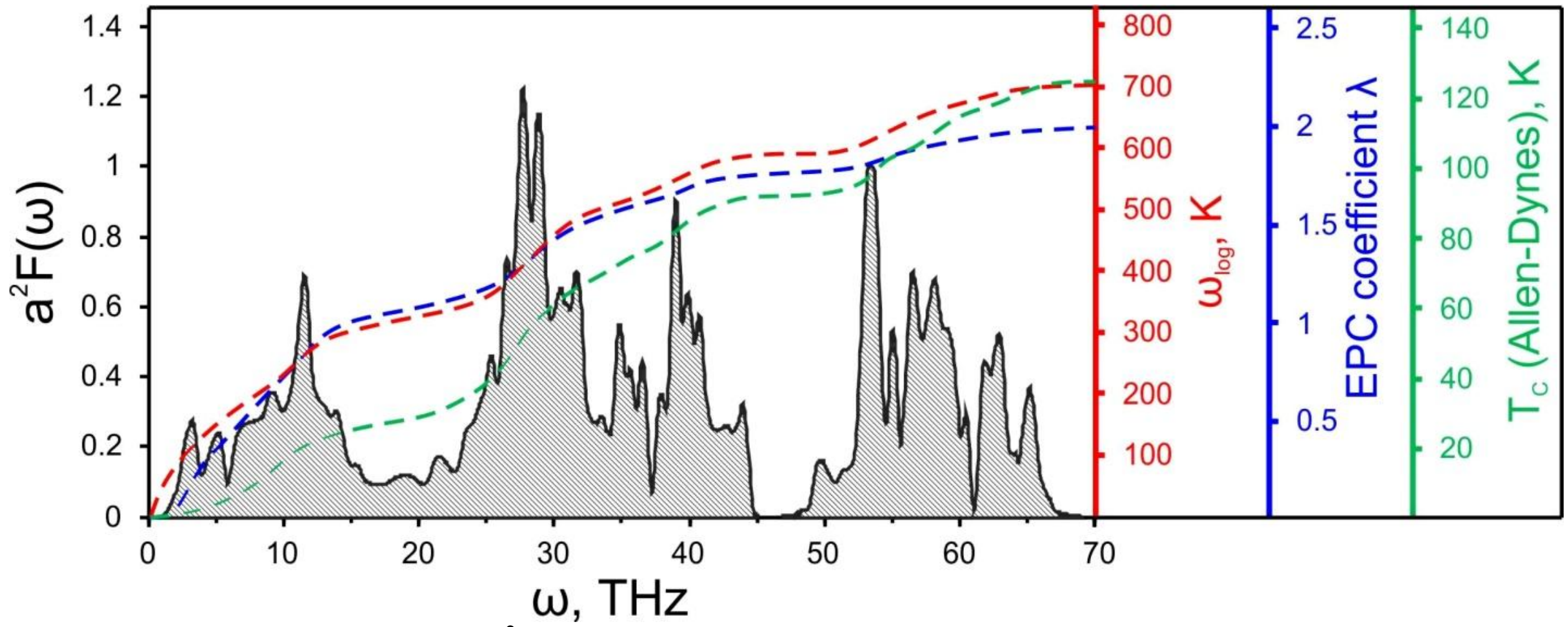

**Figure S33.** Eliashberg function $a^2F(\omega)$ of *I4/mmm*-$AcH_{12}$ at 150 GPa.

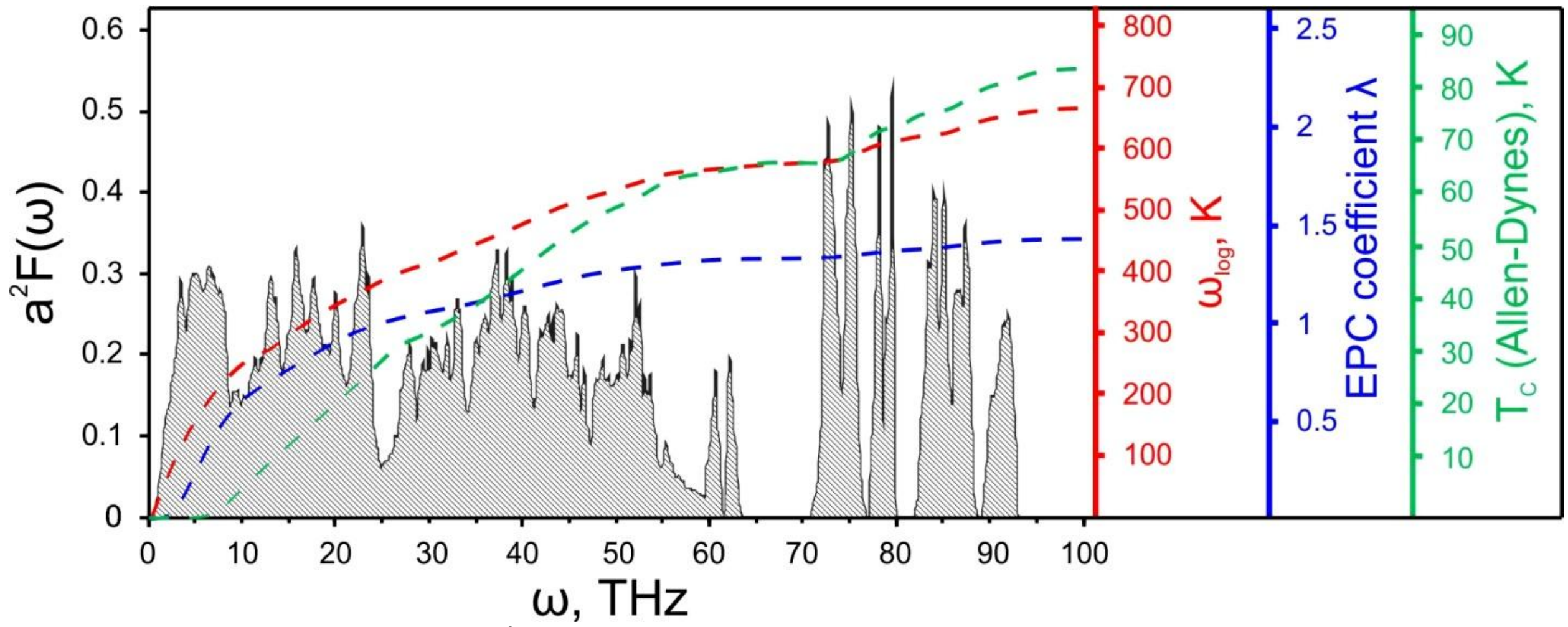

**Figure S34.** Eliashberg function $a^2F(\omega)$ of *I4/mmm*-$AcH_{12}$ at 300 GPa.

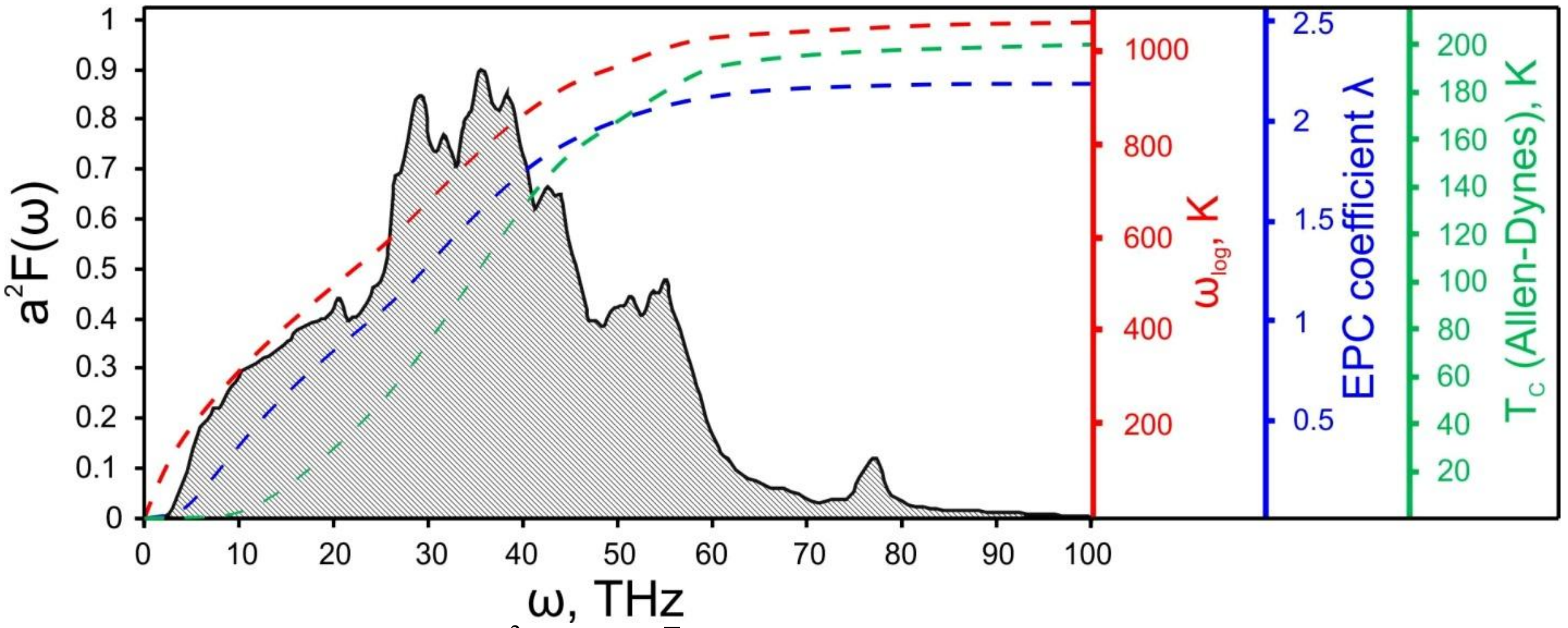

**Figure S35.** Eliashberg function $a^2F(\omega)$ of $P\bar{6}m2$-$AcH_{16}$ at 150 GPa.

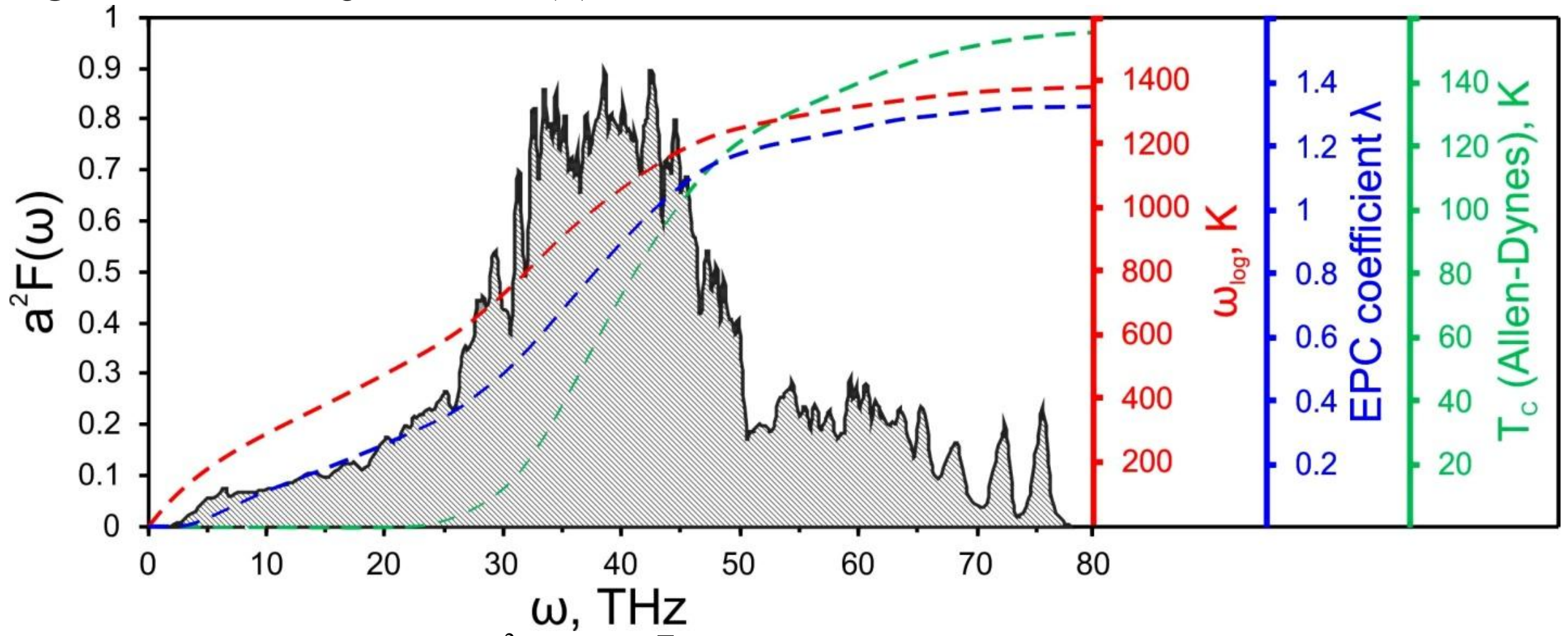

**Figure S36.** Eliashberg function $a^2F(\omega)$ of $P\bar{6}m2$-$AcH_{16}$ at 250 GPa.